\RequirePackage{fix-cm}
\documentclass[a4paper,english,aps,prd,twocolumn,10pt,a4paper,superscriptaddress,nofootinbib,nobibnotes,altaffillsymbol,showpacs,showkeys,floatfix]{revtex4-1}
\usepackage[T1]{fontenc}
\usepackage[utf8]{inputenc}
\usepackage{color}
\usepackage{babel}
\usepackage{mathrsfs}
\usepackage{amsmath}
\usepackage{amssymb}
\usepackage{graphicx}
\usepackage{esint}
\usepackage[unicode=true,
 bookmarks=true,bookmarksnumbered=false,bookmarksopen=true,bookmarksopenlevel=2,
 breaklinks=true,pdfborder={0 0 0},backref=false,colorlinks=true]
 {hyperref}
\hypersetup{pdftitle={Thirring model at finite density in 0+1 dimensions with stochastic quantization: Crosscheck with an exact solution},
 pdfauthor={Jan Pawlowski, Christian Zielinski},
 linkcolor=BlueViolet,anchorcolor=BlueViolet,citecolor=BlueViolet,filecolor=BlueViolet,urlcolor=BlueViolet}

\makeatletter

\pdfpageheight\paperheight
\pdfpagewidth\paperwidth

\newcommand{\lyxdot}{.}

 \@ifundefined{textcolor}{}
 {%
   \definecolor{BLACK}{gray}{0}
   \definecolor{WHITE}{gray}{1}
   \definecolor{RED}{rgb}{1,0,0}
   \definecolor{GREEN}{rgb}{0,1,0}
   \definecolor{BLUE}{rgb}{0,0,1}
   \definecolor{CYAN}{cmyk}{1,0,0,0}
   \definecolor{MAGENTA}{cmyk}{0,1,0,0}
   \definecolor{YELLOW}{cmyk}{0,0,1,0}
 }

\usepackage[usenames,dvipsnames]{xcolor}

\allowdisplaybreaks
\usepackage{slashed}
\usepackage{bbm}

\@ifundefined{showcaptionsetup}{}{%
 \PassOptionsToPackage{caption=false}{subfig}}
\usepackage{subfig}
\makeatother

\begin{document}

\title{Thirring model at finite density in $0+1$ dimensions\\
with stochastic quantization: Crosscheck with an exact solution}

\author{Jan M.~Pawlowski}

\affiliation{Institut für Theoretische Physik, Universität Heidelberg, Philosophenweg
16, 69120 Heidelberg, Germany}

\affiliation{ExtreMe Matter Institute EMMI, GSI, Planckstraße 1, D-64291 Darmstadt,
Germany}

\author{Christian Zielinski}

\altaffiliation{Permanent address: Division of Mathematical Sciences, Nanyang Technological University, Singapore 637371}

\affiliation{Institut für Theoretische Physik, Universität Heidelberg, Philosophenweg
16, 69120 Heidelberg, Germany}

\date{\today}
\begin{abstract}
We consider a generalized Thirring model in $0+1$ dimensions at finite
density. In order to deal with the resulting sign problem we employ
stochastic quantization, i.e., a complex Langevin evolution. We investigate
the convergence properties of this approach and check in which parameter
regions complex Langevin evolutions are applicable in this setting.
To this end we derive numerous analytical results and compare directly
with numerical results. In addition we employ indirect indicators
to check for correctness. Finally we interpret and discuss our findings. 
\end{abstract}



\maketitle

\section{Introduction}

Despite all efforts, one of the outstanding problems of lattice field
theory until this day is the sign problem. The introduction of a finite
chemical potential $\mu>0$ renders the path integral measure complex
and rapidly oscillating in many theories of interest, like quantum
chromodynamics (QCD) in $3+1$ dimensions. The oscillatory behavior
significantly increases the numerical costs, in particular in the
continuum limit. A hard sign problem exists for theories where the
costs grow more than polynomially with the volume. This obstacle hinders
numerical \textit{ab initio} studies of strongly interacting matter
under extreme conditions and the understanding of the phase diagram
of QCD. There is no satisfactory solution known to the sign problem,
despite the large number of proposed solutions.

Among the proposed solutions we can find reweighting techniques, Taylor
expansions about $\mu=0$, extrapolations from imaginary chemical
potential, the introduction of dual variables and a canonical ensemble
approach. For recent reviews see e.g.~\cite{deForcrand:2010ys,Lombardo:2005gj}.
However, due to the overlap problem reweighting techniques are computationally
expensive and can only be used for small $\mu$, while the numerical
determination of Taylor coefficients is noisy and the expansion converges
slowly \cite{Karsch:2010hm}. Also the continuation from imaginary
chemical potential is a nontrivial task \cite{Lombardo:2006yc}. The
application of dual variables and the canonical ensemble approach
is still under active research, see e.g.~\cite{Schmidt:2012uy,Delgado:2012tm}
for dual observables and \cite{Alexandru:2005ix,Alexandru:2010yb}
for simulations with canonical ensembles.

In this paper we employ a different approach. Parisi proposed already
in 1983 that stochastic quantization \cite{Parisi:1980ys}---for a
review see e.g.~\cite{Damgaard:1987rr}---could circumvent the sign
problem in terms of a complex Langevin evolution \cite{Parisi:1984cs}.
However, it is well known that the Langevin evolution may converge
towards unphysical fixed points. It has been successfully applied
to the SU(3) spin model \cite{Karsch:1985cb,Aarts:2011zn}, to an
effective theory of QCD in the strong-coupling limit \cite{Bilic:1987fn},
simple models of quantum chromodynamics \cite{Aarts:2008rr,Aarts:2010gr}
and to the relativistic Bose gas \cite{Aarts:2008wh,Aarts:2009hn}
at finite density. Furthermore it has been applied to quantum fields
in Minkowski time \cite{Berges:2006xc,Berges:2007nr}, also in nonequilibrium
\cite{Berges:2005yt}. Counterexamples are given by the three-dimensional
XY model at finite chemical potential for small $\beta$ \cite{Aarts:2010aq}
and in cases of gauge theories with static charges \cite{Ambjorn:1986fz}.
Early investigations of complex Langevin evolutions can be found in
\cite{Hamber:1985qh,Flower:1986hv,Ilgenfritz:1986cd}, while for reviews
see e.g.~\cite{Aarts:2013bla,Aarts:2009yj}. Recently, a set of consistency
conditions indicating correct convergence could be derived \cite{Aarts:2009uq,Aarts:2011ax,Aarts:2011sf}.
When truncating this infinite tower of identities one obtains necessary
conditions for correctness.

In this work we apply a complex Langevin evolution to a generalized
Thirring model at finite density. Here it serves us as a model theory
to check for the applicability of this method. Our results extend
the studies carried out in \cite{Spielmann10}, which led to ambiguous
results. In this paper we restrict ourselves to the case of $0+1$
dimensions and deal with the question of whether a complex Langevin
evolution can enable finite density calculations in this setting.
Further investigations of this approach in the $2+1$-dimensional
generalized Thirring model are presented in \cite{Pawlowski:2013gag}.
The $2+1$-dimensional model appears for example in effective theories
of high temperature superconductors and graphene, see e.g.~\cite{Gies:2010st}
and references given therein. It is also worth mentioning that in
the case of the three-dimensional massless Thirring model, a fermion
bag approach was successfully applied in \cite{Chandrasekharan:2011mn}.

We organized the paper as follows: In Sec.~\ref{sec:The-Thirring-Model}
we introduce a generalized Thirring model and its formulation on the
lattice. We discuss the Langevin equation and its numerical implementation.
In Sec.~\ref{sec:Analytical-Results} we present a closed expression
for the partition function of the lattice theory and derive some observables
of interest. We also discuss additional indicators to evaluate the
convergence properties of the complex Langevin evolution. In Sec.~\ref{sec:Numerical-Results}
we discuss the results of the numerical part of this work and aim
to answer the question in which parameter regime results are reliable.
We end this paper with concluding remarks in Sec.~\ref{sec:Conclusions}.

\global\long\def\Tr{\operatorname{Tr}}

\global\long\def\myRe{\operatorname{Re}}

\global\long\def\myIm{\operatorname{Im}}

\global\long\def\matrixOne{\mathbbm{1}}

\global\long\def\ii{\textrm{i}}

\section{The Generalized Thirring Model \label{sec:The-Thirring-Model}}

\subsection{Continuum formulation \label{sub:Continuum-formulation}}

We consider a generalization of the Thirring model. The historical
model was introduced in 1958 by Walter E.~Thirring and is one of
the rare examples of an exactly solvable quantum field theory \cite{Thirring:1958in}.
While the original model describes self-interaction fermions in $1+1$
dimensions, we consider $N_{f}$ fermion flavors at finite density.

We begin with a generalization to $d$ dimensions and then later specialize
to the case of $0+1$ dimensions. The Euclidean Lagrangian in the
continuum reads
\begin{multline}
\mathscr{L}_{\Psi}=\sum_{i=1}^{N_{f}}\overline{\Psi}_{i}\left(\slashed{\partial}+m_{i}+\mu_{i}\gamma_{0}\right)\Psi_{i}\\
+\frac{g^{2}}{2N_{f}}\left(\sum_{i=1}^{N_{f}}\overline{\Psi}_{i}\gamma_{\nu}\Psi_{i}\right)^{2}.\label{eq:TMLagrangeHist}
\end{multline}
The index $i=1,\ldots,N_{f}$ enumerates fermion flavors, $m_{i}$
and $\mu_{i}$ denote the bare mass and bare fermion chemical potential
of the respective flavor and $g^{2}$ is the bare coupling strength.
The $\gamma$ matrices satisfy the Clifford algebra $\left\{ \gamma_{\mu},\gamma_{\nu}\right\} =2\delta_{\mu\nu}\matrixOne$.

The model shows breaking of chiral symmetry at $\mu=0$ in $2+1$
dimensions \cite{Christofi:2007ye}. For the $1+1$-dimensional Thirring
model, the equivalence to the sine-Gordon model can be shown \cite{Coleman:1974bu,Benfatto:2007qp}.

The four-point interaction can be resolved with the introduction of
an auxiliary field $A_{\nu}$. This formulation reads
\begin{equation}
\mathscr{L}=\sum_{i}\overline{\Psi}_{i}\left(\slashed{\partial}+\ii\slashed{A}+m_{i}+\mu_{i}\gamma_{0}\right)\Psi_{i}+N_{f}\beta A_{\nu}^{2}.
\end{equation}
Here we introduced the inverse coupling $\beta=1/\left(2g^{2}\right)$.
When integrating $A_{\nu}$ out, we recover \eqref{eq:TMLagrangeHist}.
Although the auxiliary field $A_{\nu}$ is not a gauge field, the
model can be interpreted as a more general gauge theory after gauge
fixing, see e.g.~\cite{Itoh:1994cr}. After integrating out the fermionic
degrees of freedom we find
\begin{align}
Z & =\intop\mathscr{D}A\,\left(\prod_{i}\det K_{i}\right)\, e^{-S_{A}}=\intop\mathscr{D}A\, e^{-S_{\textrm{eff}}},\nonumber \\[2ex]
S_{A} & =N_{f}\beta\intop_{0}^{1/T}\textrm{d}t\intop\textrm{d}^{d-1}\mathbf{x}\, A_{\nu}^{2}.\label{eq:Cont-Part-Func}
\end{align}
Here we introduced the temperature $T$ and $K_{i}=\slashed{\partial}+\ii\slashed{A}+m_{i}+\mu_{i}\gamma_{0}$.
Including the fermion determinant in the exponential term yields
\begin{equation}
S_{\textrm{eff}}=S_{A}-\sum_{i}\Tr\log K_{i}.
\end{equation}
For the fermion determinant the relation
\begin{equation}
\det K_{i}\left(\mu\right)=\left[\det K_{i}\left(-\mu^{\star}\right)\right]^{\star}\label{eq:Fermion-det-symmetry}
\end{equation}
holds, thus rendering the path integral measure complex for $\mu>0$.
At vanishing or purely imaginary chemical potential, the determinant
is real and the theory is free of a sign problem. If the fermion determinant
is replaced by its modulus, we refer to this as the phase-quenched
case. Physically this corresponds to the introduction of an isospin
chemical potential.

Like quantum chromodynamics, the Thirring model exhibits Silver Blaze
behavior \cite{Cohen:2003kd,Splittorff:2006fu}. It implies that at
vanishing temperature there is a threshold $\mu_{c}$, so that observables
are independent of the chemical potential $\mu$ for $\mu<\mu_{c}$.
While in the full theory the onset is given by the physical fermion
mass $m_{\textrm{phys}}$, in the phase-quenched theory we have $\mu_{c}=m_{\pi}/2$,
where $m_{\pi}$ is the physical pion mass.

\subsection{Lattice formulation}

We consider the case of $0+1$ dimensions---corresponding to a quantum
mechanical system---with lattice spacing $a$ and $N_{t}$ lattice
points. We employ staggered fermions \cite{Kogut:1974ag,Banks:1975gq,Banks:1976ia,Susskind:1976jm}
and denote the number of lattice flavors, i.e., the number of staggered
fermion fields, by $\mathcal{N}$. Furthermore we assume that $N_{t}$
is even, as otherwise the formulation of staggered fermions is conceptually
problematic and \eqref{eq:Fermion-det-symmetry} is violated. In order
to introduce a finite chemical potential $\mu$, we use the prescription
by Hasenfratz and Karsch \cite{Hasenfratz:1983ba}. For notational
ease we refer to the one-component auxiliary field as $A_{t}=A_{0}(x=t)$
and all dimensionful quantities are scaled dimensionless by appropriate
powers of $a$. Furthermore we introduce the hopping parameter $\kappa=1/\left(2m\right)$.
The temperature corresponds to the inverse temporal extension $T=N_{t}^{-1}$.

Using this formulation the lattice partition function reads
\begin{equation}
Z=\intop_{-\infty}^{\infty}\prod_{t=1}^{N_{t}}\textrm{d}A_{t}\,\left(\prod_{i}\det K_{i}\right)\, e^{-S_{A}}\label{eq:Latt-Part-Func}
\end{equation}
with $S_{A}=\frac{1}{2}\mathcal{N}\beta\sum_{t}A_{t}^{2}$ and flavor
index $i=1,\ldots,\mathcal{N}$. The fermion matrix takes the form
\begin{multline}
K_{i}\left(t,\tau\right)=\frac{1}{2}\left(1+\ii A_{t}\right)e^{\mu_{i}}\delta_{t+1,\tau}\\
-\frac{1}{2}\left(1-\ii A_{\tau}\right)e^{-\mu_{i}}\delta_{t-1,\tau}+m_{i}\delta_{t\tau},\label{eq:Fermion-matrix}
\end{multline}
where we impose antiperiodic boundary conditions, cf.~\cite{DelDebbio:1995zc,Spielmann10}.
In our analysis we focus on a few observables, namely the fermion
density and condensate, the energy density and the phase factor of
the fermion determinant. In the following, sums over the flavor index
$i$ are not implied. The fermion density of a given flavor is given
by
\begin{equation}
\left\langle n_{i}\right\rangle =\frac{1}{N_{t}}\left(\frac{\partial\log Z}{\partial\mu_{i}}\right)_{T}=\frac{1}{N_{t}}\left\langle \Tr\left(\frac{\partial K_{i}}{\partial\mu_{i}}K_{i}^{-1}\right)\right\rangle .
\end{equation}
The fermion condensate follows from
\begin{equation}
\left\langle \overline{\chi}_{i}\chi_{i}\right\rangle =\frac{1}{N_{t}}\left(\frac{\partial\log Z}{\partial m_{i}}\right)_{T,\mu_{i}}=\frac{1}{N_{t}}\left\langle \Tr K_{i}^{-1}\right\rangle 
\end{equation}
and the energy density reads
\begin{equation}
\left\langle \varepsilon_{i}\right\rangle =-\left(\frac{\partial\log Z}{\partial N_{t}}\right)_{\mu_{i}}+\mu_{i}\left\langle n_{i}\right\rangle ,
\end{equation}
which we normalize to $\left\langle \varepsilon_{i}\right\rangle \left(\mu=0\right)=0$.

The phase factor of the determinant is defined by $\exp\left(\ii\phi\right)=\det K/\left|\det K\right|$.
It can be expressed in terms of the partition function
\begin{equation}
Z_{\mathcal{N}}=\intop_{-\infty}^{\infty}\prod_{t}\textrm{d}A_{t}\,\left(\det K\right)^{\mathcal{N}}\, e^{-S_{A}},\label{eq:Latt-Part-Func-N}
\end{equation}
for $\mathcal{N}$ degenerated flavors and $Z_{\mathcal{N}}^{\textrm{pq}}$
for the phase-quenched case, where the fermion determinant in \eqref{eq:Latt-Part-Func-N}
is replaced by its modulus. The expectation value of $\exp\left(\ii\mathcal{N}\phi\right)$
follows in the $\mathcal{N}$ flavor phase-quenched theory \cite{Han:2008xj,Andersen:2009zm}
as
\begin{equation}
\left\langle e^{\ii\mathcal{N}\phi}\right\rangle _{\mathcal{N}}^{\textrm{pq}}=\frac{Z_{\mathcal{N}}}{Z_{\mathcal{N}}^{\textrm{pq}}}\in\left[0,1\right].
\end{equation}
A value close to zero indicates a rapidly oscillating path integral
measure with a severe sign problem.

\subsection{Complex Langevin evolution \label{sub:Complex-Langevin-evolution}}

The idea of stochastic quantization is that observables in a Euclidean
quantum field theory can be obtained as the equilibrium values of
a statistical system coupled to a heat bath \cite{Damgaard:1987rr}.
The problem of quantizing a field theory is then reduced to finding
the static solutions of an associated Langevin equation. If the action
is real and bounded from below, correctness of this approach can be
ensured. We can also formally generalize to the case of a complex
action \cite{Parisi:1984cs}. This situation naturally arises when
considering field theories at finite density. Until this day there
is a lack of rigor mathematical understanding regarding the validity
of this procedure. However, in cases where it is converging correctly
one has a very elegant solution for the sign problem at hand.

We aim to check for the applicability of complex Langevin evolutions
to the Thirring model. To this end we have to find the static solution
of the Langevin equation
\begin{equation}
\frac{\partial}{\partial\Theta}A_{t}\left(\Theta\right)=-\frac{\delta S_{\textrm{eff}}\left[A\right]}{\delta A_{t}\left(\Theta\right)}+\sqrt{2}\,\eta_{t}\left(\Theta\right),\label{eq:Cont-Langevin-eq}
\end{equation}
where $\Theta$ denotes a fictitious time. The noise term $\eta_{t}\left(\Theta\right)$
follows a Gaussian distribution with
\begin{align}
\left\langle \eta_{t}\left(\Theta\right)\right\rangle  & =0,\nonumber \\
\left\langle \eta_{t}\left(\Theta\right)\eta_{t'}\left(\Theta'\right)\right\rangle  & =\delta\left(t-t'\right)\delta\left(\Theta-\Theta'\right).
\end{align}
A simple approach to solve the Langevin equation numerically is a
first order integration scheme with fixed stepsize $\epsilon_{L}$.
Higher order integration schemes of $\mathcal{O}(\epsilon_{L}^{3/2})$
have been employed in the literature too \cite{Chang:1987,Aarts:2011zn}.
However, in some models fixed stepsize integration schemes fail due
to the occurrence of run-away trajectories, which can be avoided by
the use of an adaptive stepsize \cite{Ambjorn:1985iw,Aarts:2009dg}.
Although a constant stepsize proved here to be sufficient \cite{Spielmann10},
we employ an adaptive stepsize algorithm due to better convergence
properties. For $\mathcal{N}$ degenerated flavors our discretization
of \eqref{eq:Cont-Langevin-eq} reads
\begin{equation}
A_{t}\left(\Theta+\epsilon_{L}\right)=A_{t}\left(\Theta\right)+\epsilon_{L}D_{t}\left(\Theta\right)+\sqrt{2\epsilon_{L}}\,\eta_{t}\left(\Theta\right)\label{eq:Latt-Langevin-eq}
\end{equation}
with drift term
\begin{multline}
D_{t}\left(\Theta\right)=-\mathcal{N}\beta A_{t}\left(\Theta\right)\\
+\frac{\mathcal{N}\ii}{2}\left[K^{-1}\left(t+1,t\right)e^{\mu}+K^{-1}\left(t,t+1\right)e^{-\mu}\right].\label{eq:Drift-term}
\end{multline}
After each integration step the stepsize $\epsilon_{L}$ will be updated
according to
\begin{equation}
\epsilon_{L}\equiv\epsilon_{L}\left(\Theta\right)=\frac{\delta}{\max_{t}\left|D_{t}\left(\Theta\right)\right|}\label{eq:Adapative-stepsize}
\end{equation}
with stepsize parameter $\delta=10^{-3}$ (compare to \cite{Spielmann10}).

It is possible to generalize the real noise term in \eqref{eq:Latt-Langevin-eq}
to an imaginary one \cite{Spielmann10} via the replacement
\begin{equation}
\eta_{t}\left(\Theta\right)\;\to\;\sqrt{\mathcal{I}+1}\myRe\eta_{t}\left(\Theta\right)+\ii\sqrt{\mathcal{I}}\myIm\eta_{t}\left(\Theta\right)
\end{equation}
with $\mathcal{I}\geq0$. The noise correlators then read
\begin{align}
\left\langle \myRe\eta_{t}\left(\Theta\right)\myRe\eta_{t'}\left(\Theta'\right)\right\rangle  & =\left\langle \myIm\eta_{t}\left(\Theta\right)\myIm\eta_{t'}\left(\Theta'\right)\right\rangle \nonumber \\
 & =\delta\left(t-t'\right)\delta\left(\Theta-\Theta'\right)
\end{align}
and $\left\langle \myRe\eta_{t}\left(\Theta\right)\myIm\eta_{t'}\left(\Theta'\right)\right\rangle =0$.
Assuming correctness of the complex Langevin evolution and numerical
stability, we expect expectation values to be independent of $\mathcal{I}$.

\section{Analytical Results \label{sec:Analytical-Results}}

\subsection{Exact partition function}

\begin{figure}
\subfloat[Plot of the fermion density $\left\langle n\right\rangle $.]{\includegraphics[width=1\columnwidth]{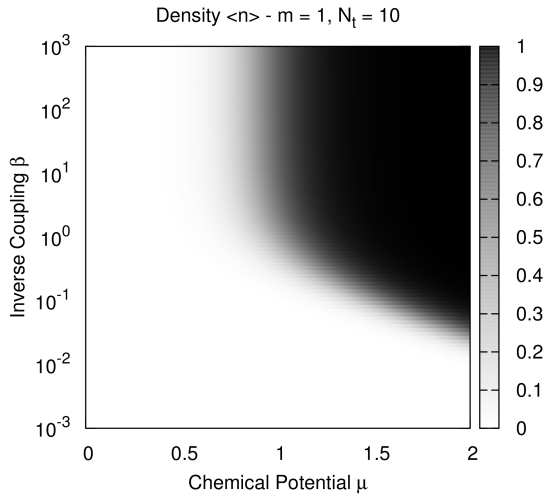}

}

\subfloat[Plot of the fermion condensate $\left\langle \overline{\chi}\chi\right\rangle $.]{\includegraphics[width=1\columnwidth]{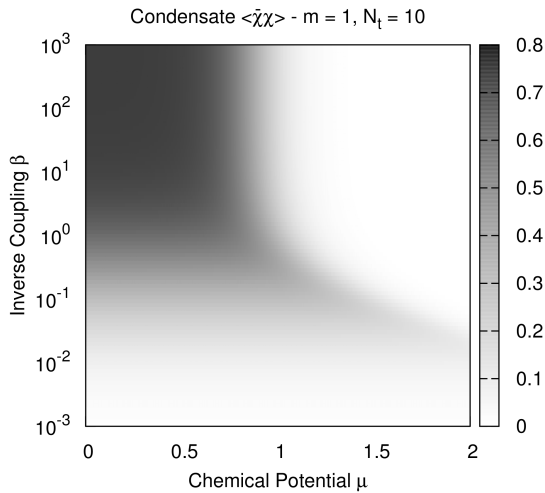}

}

\caption{In the phase structure in $d=0+1$ we find a condensed phase for large
$\mu$. \label{fig:Phase-diagram}}
\end{figure}

We begin with the partition function for one staggered fermion field,
i.e., $\mathcal{N}=1$. We incorporate antiperiodic boundary conditions
and for brevity we introduce
\begin{align}
\mathfrak{B}_{\pm} & =\frac{1}{2\left(2\kappa\right)^{N_{t}}}\left(1\pm\sqrt{\mathfrak{B}_{c}/\beta}\right)^{N_{t}},\nonumber \\
\mathfrak{B}_{c} & =\beta+4\left(\beta+1\right)\kappa^{2}.
\end{align}
Then the partition function \eqref{eq:Latt-Part-Func} reads
\begin{equation}
Z_{1}=2\left(\frac{\pi}{2\beta}\right)^{N_{t}/2}\left[\mathfrak{B}_{+}+\mathfrak{B}_{-}+\cosh\left(N_{t}\mu\right)\right].\label{eq:Latt-Exact-Part-Func}
\end{equation}
This can be shown for example by systematic saturation of the Grassmann
integral or the help of the determinant identities in \cite{Molinari20082221}.
For the fermion density we find
\begin{equation}
\left\langle n\right\rangle =\frac{\sinh\left(\mu/T\right)}{\mathfrak{B}_{+}+\mathfrak{B}_{-}+\cosh\left(\mu/T\right)},
\end{equation}
while the fermion condensate is given by
\begin{equation}
\left\langle \overline{\chi}\chi\right\rangle =\frac{2\kappa\sqrt{\beta/\mathfrak{B}_{c}}\left(\mathfrak{B}_{+}-\mathfrak{B}_{-}\right)}{\mathfrak{B}_{+}+\mathfrak{B}_{-}+\cosh\left(\mu/T\right)}.
\end{equation}
The expression for the energy density is rather lengthy and we will
not quote it here explicitly. Figure \ref{fig:Phase-diagram} shows
the dependence of these observables on both $\beta$ and $\mu$. For
large $\mu$ we find a condensed phase, which is well separated for
large $N_{t}$.

As it turns out, we can take the continuum limit of the density $\left\langle n\right\rangle $
analytically. To this end we recover the physical units of all dimensionful
quantities by reintroducing the lattice spacing $a$. We fix the dimensionful
temperature $T^{-1}=aN_{t}$, express the lattice spacing $a$ as
a function of the number of lattice points $N_{t}$ and take the limit
$N_{t}\to\infty$. We obtain
\begin{equation}
\left\langle n\right\rangle _{\textrm{cont}}=\frac{\sinh\left(\frac{\mu}{T}\right)}{\frac{1}{2}\exp\left(\frac{\kappa-\beta}{2T\beta\kappa}\right)\left[1+\exp\left(\frac{1}{T\kappa}\right)\right]+\cosh\left(\frac{\mu}{T}\right)},
\end{equation}
where all units are explicitly dimensionful. In the zero temperature
limit $T\to0$ we find $\left\langle n\right\rangle _{\textrm{cont}}=\Theta\left(\mu-m_{\textrm{phys}}\right)$
with physical fermion mass $m_{\textrm{phys}}=m+g^{2}$ and bare mass
$m$.

\subsection{Several flavors \label{sub:Several-flavor-theory}}

We also considered the case of more than one lattice flavor, but only
quote here the simplest case of \eqref{eq:Latt-Part-Func-N} for $N_{t}=2$
lattice points and $\mathcal{N}=2$ staggered fermion fields
\begin{multline}
Z_{2}=\frac{\pi}{32\beta^{3}\kappa^{4}}\left[2\beta^{2}+4\beta\kappa^{2}+8\beta^{2}\kappa^{2}+5\kappa^{4}\right.\\
+12\beta\kappa^{4}+12\beta^{2}\kappa^{4}+8\beta^{2}\kappa^{2}\cosh\left(2\mu\right)\\
+\kappa^{4}\cosh\left(4\mu\right)+8\beta\kappa^{4}\cosh\left(2\mu\right)-4\beta\kappa^{4}\cosh\left(4\mu\right)\\
\left.+16\beta^{2}\kappa^{4}\cosh\left(2\mu\right)+4\beta^{2}\kappa^{4}\cosh\left(4\mu\right)\right].
\end{multline}
We observe that for two flavors the density, condensate and energy
density have plateaus in the range of $\beta\approx0.3\text{--}1.2$,
see also Fig.~\ref{fig:Multiflavor}. They appear between the onset
to the condensed phase and saturation of the density. The height of
the plateau depends on the masses of the flavors, and in the case
of degenerated flavors it is exactly at half of the maximum value
of $\left\langle n\right\rangle $ or $\left\langle \overline{\chi}\chi\right\rangle $,
respectively. If $\beta$ is very small or large, the plateaus eventually
disappear. For large $\beta$ this can be understood from the weak
coupling limit, see Sec.~\ref{sub:Weak-coupling-limit}. The existence
of these plateaus on the lattice is further confirmed by a heavy dense
limit \cite{Pawlowski:2013gag} and the Monte-Carlo studies in Sec.~\ref{sub:Multi-flavor-numerical}.
In the general case of $\mathcal{N}$ flavors, we can find up to $\mathcal{N}-1$
intermediate plateaus in these observables. We give a natural explanation
of these structures in the Appendix.

\subsection{Weak coupling limit \label{sub:Weak-coupling-limit}}

In the limit of $\beta\gg1$ the path integral measure has a strong
peak at the origin and can be approximated by a Dirac $\delta$ function.
For $\mathcal{N}$ flavors we find
\begin{multline}
\mathcal{Z}_{1}^{\textrm{weak}}=\left(\frac{2\pi}{\mathcal{N}\beta}\right)^{\frac{N_{t}}{2}}\left(\frac{2}{2^{N_{t}}}\right)^{\mathcal{N}}\\
\times\prod_{i}\left[\mathcal{B}_{i}^{+}+\mathcal{B}_{i}^{-}+\cosh\left(N_{t}\mu\right)\right]
\end{multline}
for the partition function in the weak coupling limit. Here we introduced
\begin{equation}
\mathcal{B}_{i}^{\pm}=\frac{1}{2\left(2\kappa_{i}\right)^{N_{t}}}\left(1\pm\sqrt{1+4\kappa_{i}^{2}}\right)^{N_{t}}.
\end{equation}
This can be shown again using the identities in \cite{Molinari20082221}.
Note that in this limit the contribution from different flavors factorize
and the phase-quenched case equals the full theory. For $\mathcal{N}$
degenerated flavors, i.e., $\mathcal{B}_{i}^{\pm}=\mathcal{B}_{\pm}$,
the total fermion density is given by
\begin{equation}
\left\langle n\right\rangle =\frac{\mathcal{N}\sinh\left(\mu/T\right)}{\mathcal{B}_{+}+\mathcal{B}_{-}+\cosh\left(\mu/T\right)}
\end{equation}
and the fermion condensate by
\begin{equation}
\left\langle \overline{\chi}\chi\right\rangle =\frac{2\kappa\,\mathcal{N}\left(\mathcal{B}_{+}-\mathcal{B}_{-}\right)}{\sqrt{1+4\kappa^{2}}\left(\mathcal{B}_{+}+\mathcal{B}_{-}+\cosh\left(\mu/T\right)\right)}.
\end{equation}
While approaching the noninteracting limit, the plateau structures
observed for $\mathcal{N}>1$ disappear.

\subsection{Analyticity in $\mu^{2}$ \label{sub:Analiticity-in-mu2}}

An observable $\mathcal{O}$ which is even in $\mu$ can be interpreted
as a function of $\mu^{2}$. Assuming analyticity in $\mu^{2}$, we
can analytically continue $\mathcal{O}$ to purely imaginary chemical
potential, i.e., $\mu^{2}\leq0$. In this case the fermion determinant
is real and free of a sign problem due to \eqref{eq:Fermion-det-symmetry}
and we can employ a real Langevin evolution. For $\mu^{2}>0$ we employ
a complex Langevin evolution.

Assuming the correctness of the complex Langevin evolution, $\mathcal{O}$
should be analytic at $\mu^{2}=0$. Any nonanalyticity would be an
indicator for incorrect convergence. This criterion was previously
employed in \cite{Aarts:2011zn} and in this work we apply it to the
condensate $\left\langle \overline{\chi}\chi\right\rangle $.

\subsection{Consistency conditions \label{sub:Consistency-conditions-theory}}

In \cite{Aarts:2011sf} the authors derived a set of identities indicating
correct convergence of expectation values obtained by a complex Langevin
evolution. These consistency conditions state that for all entire
holomorphic observables $\mathcal{O}$ the expectation value $\left\langle L\mathcal{O}\right\rangle =0$
has to vanish. Here
\begin{equation}
L_{t}=\left(\frac{\textrm{d}}{\textrm{d}A_{t}}+D_{t}\right)\frac{\textrm{d}}{\textrm{d}A_{t}}
\end{equation}
denotes the Langevin operator and $D_{t}=-\textrm{d}S_{\textrm{eff}}/\textrm{d}A_{t}$
the drift term.

As this defines an uncountable number of identities, we restrict the
analysis to a finite subset. We follow \cite{Aarts:2011sf} and consider
here observables $\mathcal{O}\left(t,k\right)=\exp\left(\ii kA_{t}\right)$.
The resulting conditions take the form
\begin{equation}
\left\langle L_{t}\mathcal{O}\left(t,k\right)\right\rangle =\left\langle \ii k\left[\ii k+D_{t}\right]e^{\ii kA_{t}}\right\rangle =0,
\end{equation}
which have to hold for $\forall t$ and $\forall k$. Without loss
of generality we set $t=1$. Besides $\mathcal{O}\left(t,k\right)$,
we also check the consistency conditions for the fermion density and
the condensate.

\subsection{Propagator at finite density }

\begin{figure}
\includegraphics[width=1\columnwidth]{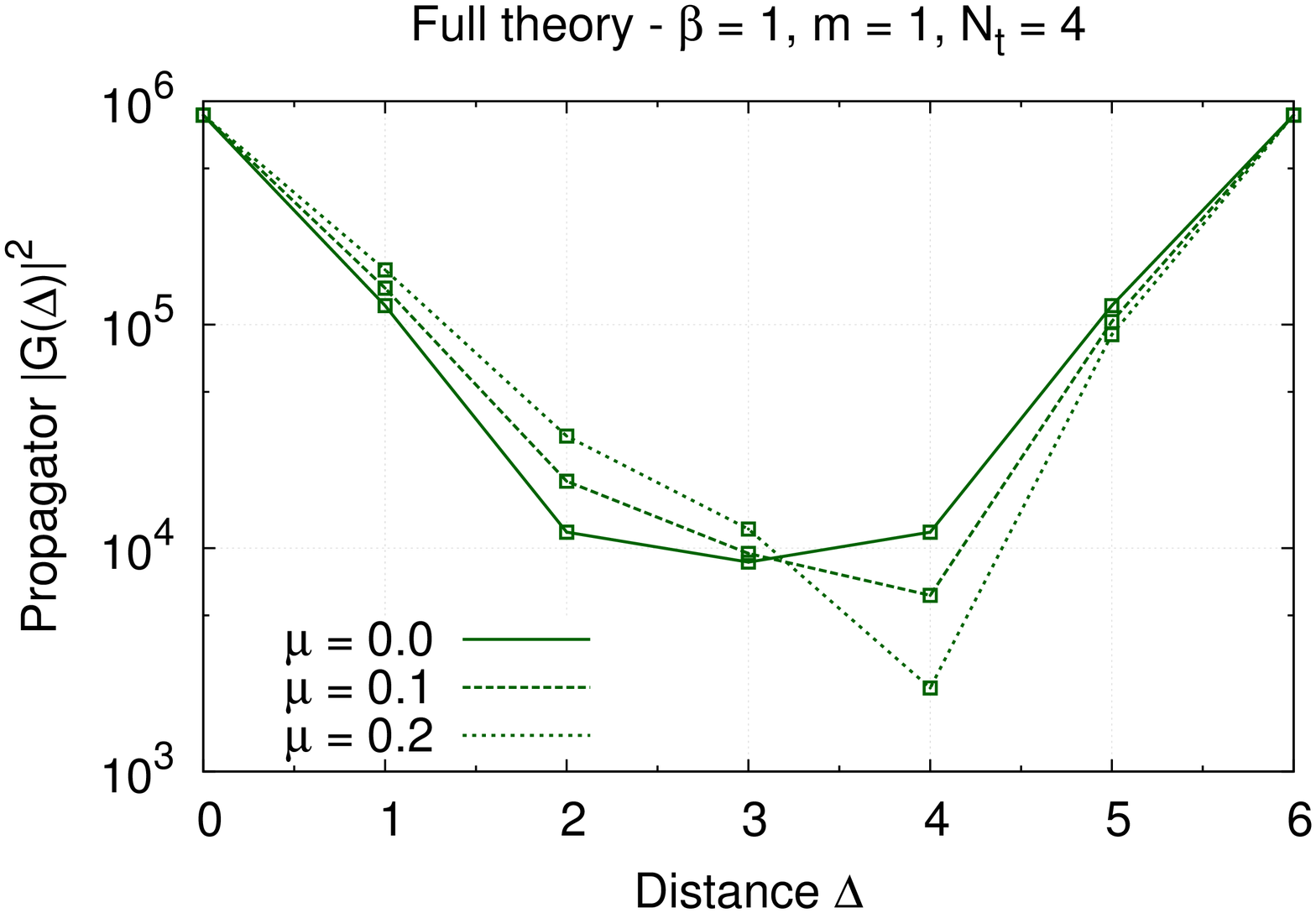}

\caption{The propagator at different values of $\mu$. \label{fig:Propagator}}
\end{figure}

We define the propagator at finite temperature $T=N_{t}^{-1}$ and
chemical potential $\mu$ by
\begin{equation}
\left\langle \chi\left(t_{1}\right)\overline{\chi}\left(t_{2}\right)\right\rangle =\frac{1}{Z_{1}}\left\langle K^{-1}\left(t_{1},t_{2}\right)\right\rangle 
\end{equation}
with partition function $Z_{1}$ given by \eqref{eq:Latt-Exact-Part-Func}.
It is helpful to introduce the notation
\begin{equation}
G\left(\Delta\right)=\left\langle \chi\left(1\right)\overline{\chi}\left(1+\hat{\Delta}\right)\right\rangle 
\end{equation}
with $\hat{\Delta}=\Delta\,\textrm{mod}\, N_{t}$. For small lattices,
the inversion of the fermion matrix and the calculation of the expectation
value can be done analytically. A typical example can be found in
Fig.~\ref{fig:Propagator}. For small distances the propagator $G\left(\Delta\right)$
falls off exponentially, but due to the finite lattice extension eventually
rises again. The introduction of a finite chemical potential $\mu$
shifts the minimum and results in $G\left(\Delta\right)\neq G\left(-\Delta\right)$.

\section{Numerical Results \label{sec:Numerical-Results}}

\subsection{Implementation}

\begin{figure}
\includegraphics[width=1\columnwidth]{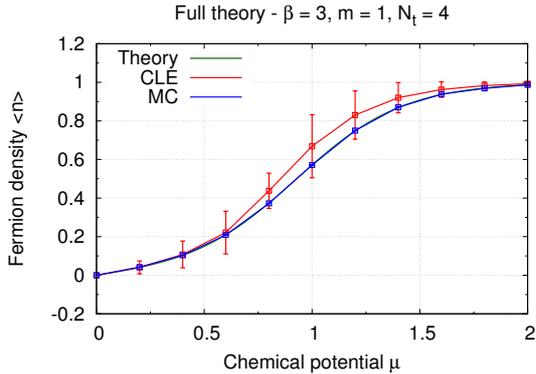}

\caption{Typical errors estimated by the standard deviation. \label{fig:Bootstrap-vs-std-dev}}
\end{figure}

In the previous section we derived numerous analytical results, which
allow us to benchmark the complex Langevin evolution and check for
its correctness. For the numerical simulations we implemented an adaptive
numerical integration scheme of the associated Langevin equation as
described in Sec.~\ref{sub:Complex-Langevin-evolution}. Using a
computer algebra system, we can determine the inverse of the fermion
matrix in \eqref{eq:Fermion-matrix} and find an analytical expression
of the drift term in \eqref{eq:Drift-term} for small lattices in
order to minimize numerical errors.

The evaluation of the observables for a given set of parameters begins
with a hot start, i.e., the random initialization of the auxiliary
field, followed by typically $10^{4}$ steps for thermalization. The
field configuration is then sampled about $\mathcal{O}\left(10^{5}\right)$
times and used to evaluate the observables. To reduce potential autocorrelation
effects, two subsequent samples are separated by ten dummy steps.
In all following plots, numerical values obtained by a complex Langevin
evolution will be denoted by ``CLE,'' while analytical results are
denoted by ``Theory.''

A simple method to estimate the error is to take the standard deviation
of the different samples of a given observable in a particular run.
However, the resulting errors are much larger than the empirical observed
statistical fluctuations between different runs and are overestimated,
see Fig.~\ref{fig:Bootstrap-vs-std-dev} for a typical example. In
order to obtain more reliable error bounds, we employ a bootstrap
analysis \cite{DeGrand:2006zz}. Besides the statistical error, we
also have to deal with systematic errors induced by a finite stepsize.
To this end we checked that the stepsize parameter $\delta$ in \eqref{eq:Adapative-stepsize}
is sufficiently small.

As an additional test we used an adaptive quasi-Monte Carlo strategy
\cite{Morokoff1995218} for the direct evaluation of expectation values,
which works well for sufficiently small $N_{t}$ and $\mu$. On larger
lattices the sign problem is severe and the algorithm fails. As the
path integral measure falls off rapidly for large field configurations,
we replace the numerically difficult noncompact integration over the
auxiliary field by a compact domain $\left[-\Lambda,\Lambda\right]^{N_{t}}$.
We parametrize this \textit{ad hoc} cutoff $\Lambda$ by
\begin{equation}
\Lambda=\sqrt{\frac{2\sigma}{\mathcal{N}\beta}}.
\end{equation}
If the geometric mean of the field configuration is $\Lambda$, then
the path integral measure is dampened by a factor of $\exp\left(-\sigma\right)$.
If we choose $\sigma$ too small, we introduce a large cutoff effect.
If $\sigma$ is very large, we still have the problem that the integrand
is close to zero in most of the integration region. As a compromise
here we use $\sigma=10$. In figures, we refer to numerical results
obtained by this method as ``MC.''

\subsection{Comparison of results}

\begin{figure}[p]
\subfloat[Fermion density $\left\langle n\right\rangle $.]{\includegraphics[width=1\columnwidth]{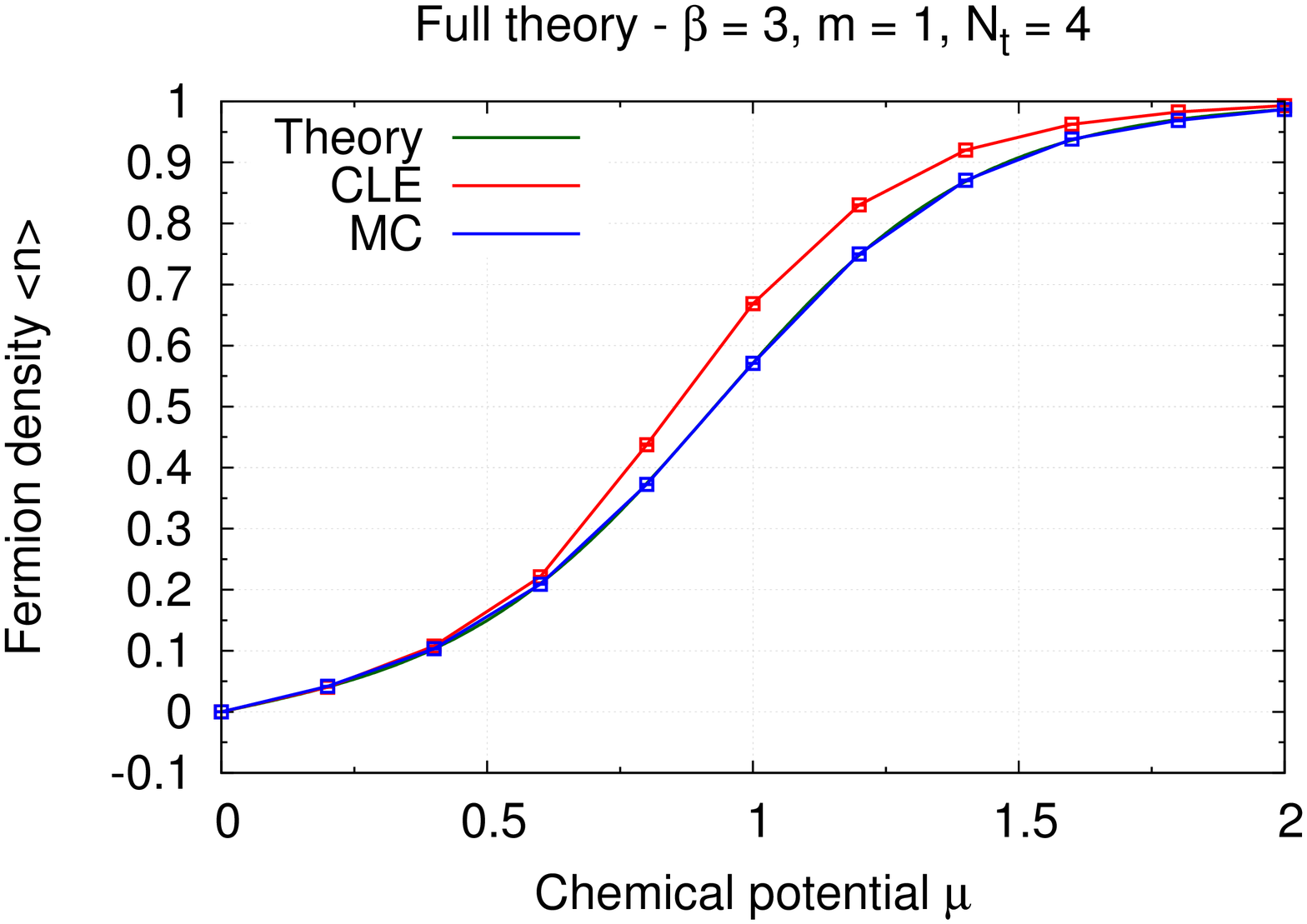}

}

\subfloat[Fermion condensate $\left\langle \overline{\chi}\chi\right\rangle $.]{\includegraphics[width=1\columnwidth]{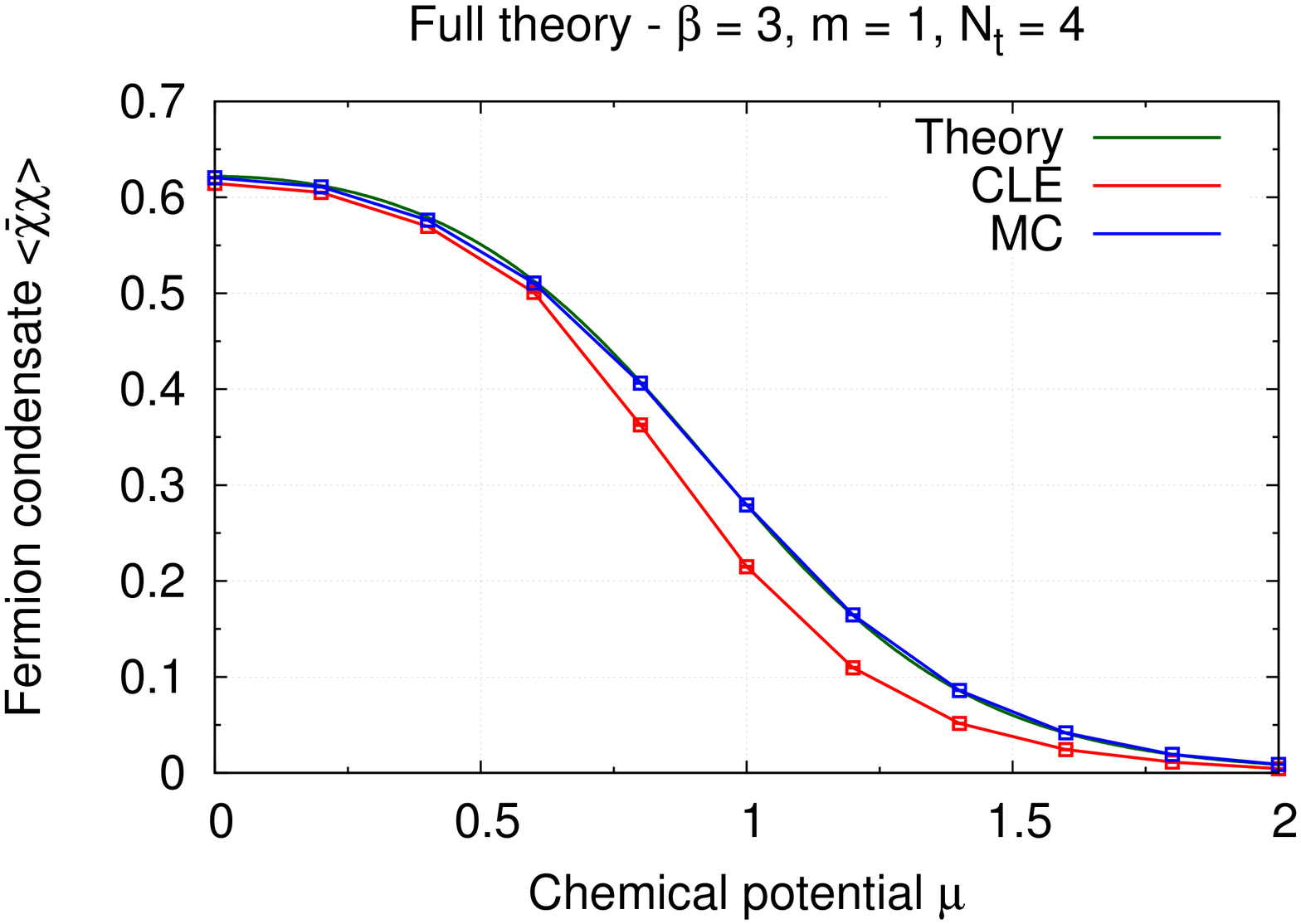}

}

\subfloat[Phase factor of determinant $\left\langle e^{\ii\phi}\right\rangle $.]{\includegraphics[width=1\columnwidth]{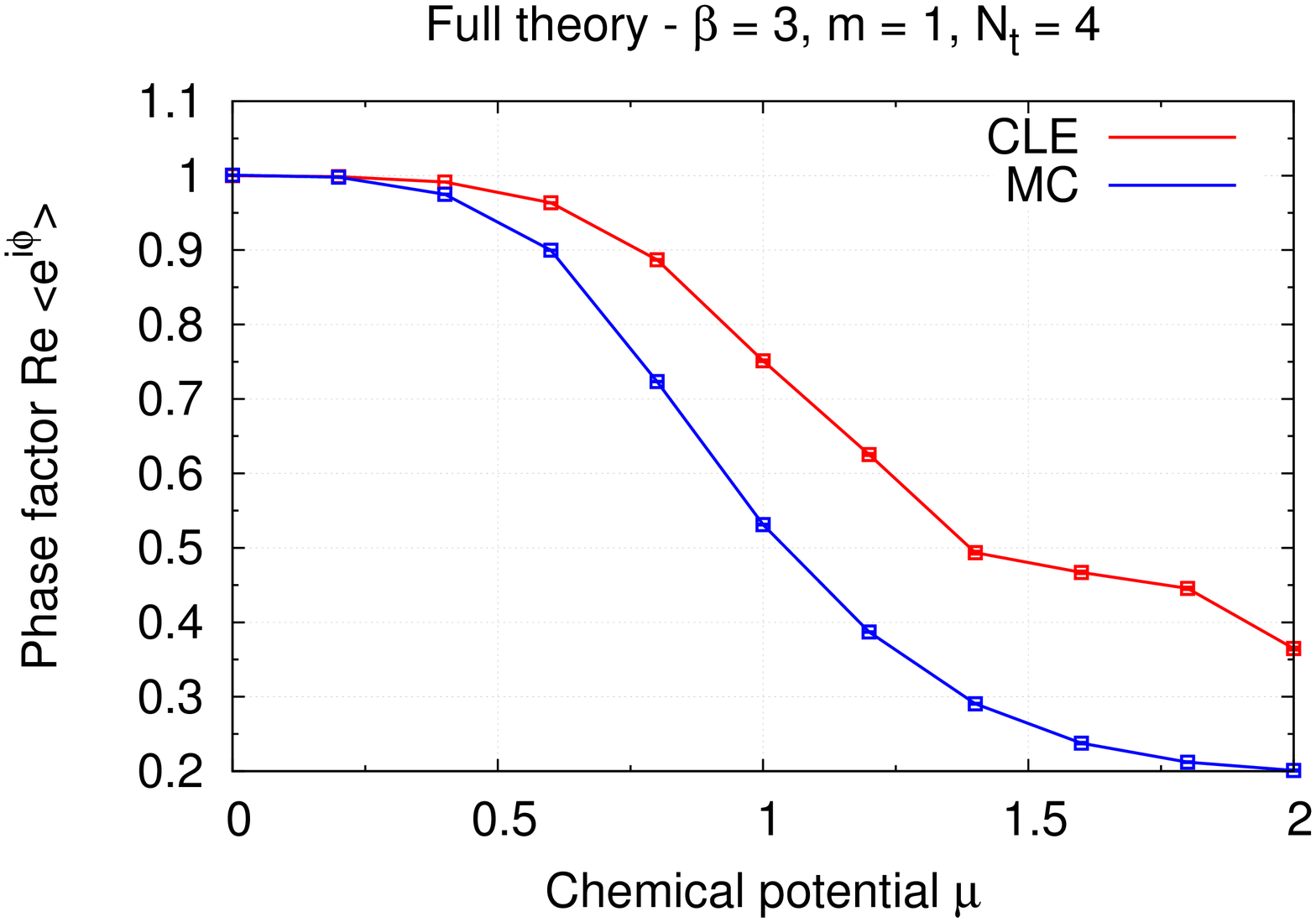}

}

\caption{Benchmarking results for $\beta=3$. \label{fig:Comparision-of-results-4-3-1} }
\end{figure}
\begin{figure}
\subfloat[Fermion density $\left\langle n\right\rangle $.]{\includegraphics[width=1\columnwidth]{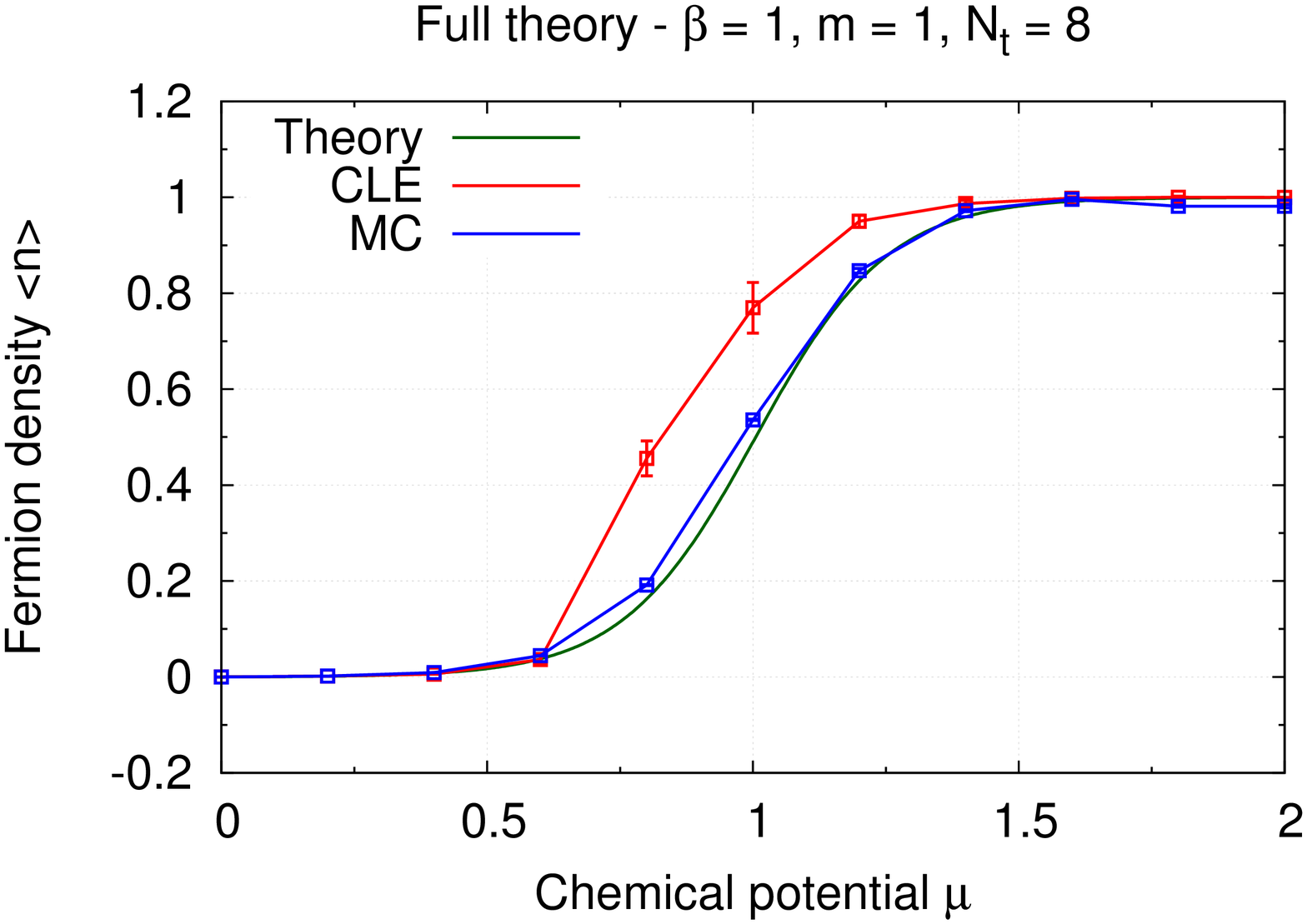}

}

\subfloat[Fermion condensate $\left\langle \overline{\chi}\chi\right\rangle $.]{\includegraphics[width=1\columnwidth]{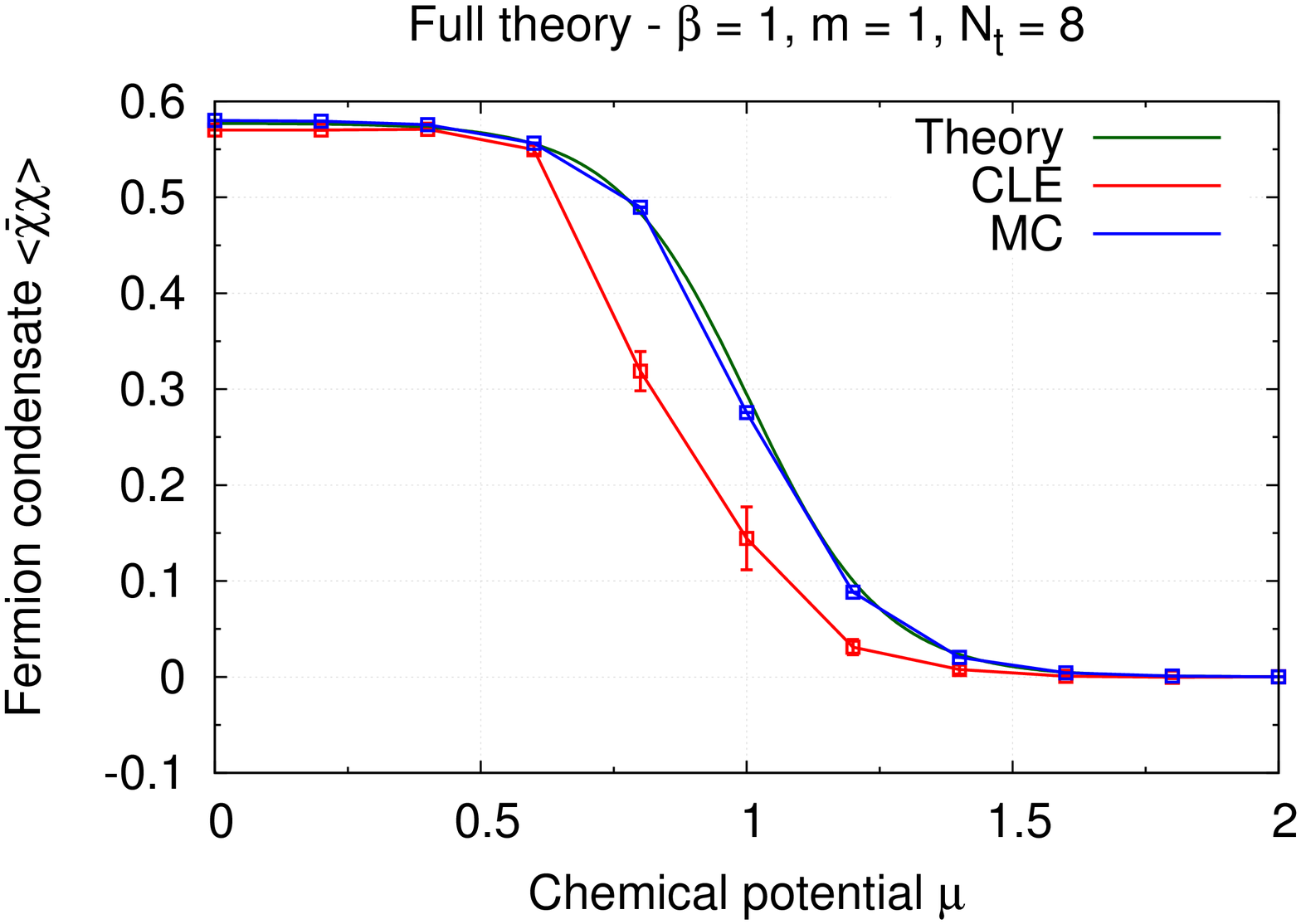}

}

\subfloat[Phase factor of determinant $\left\langle e^{\ii\phi}\right\rangle $.]{\includegraphics[width=1\columnwidth]{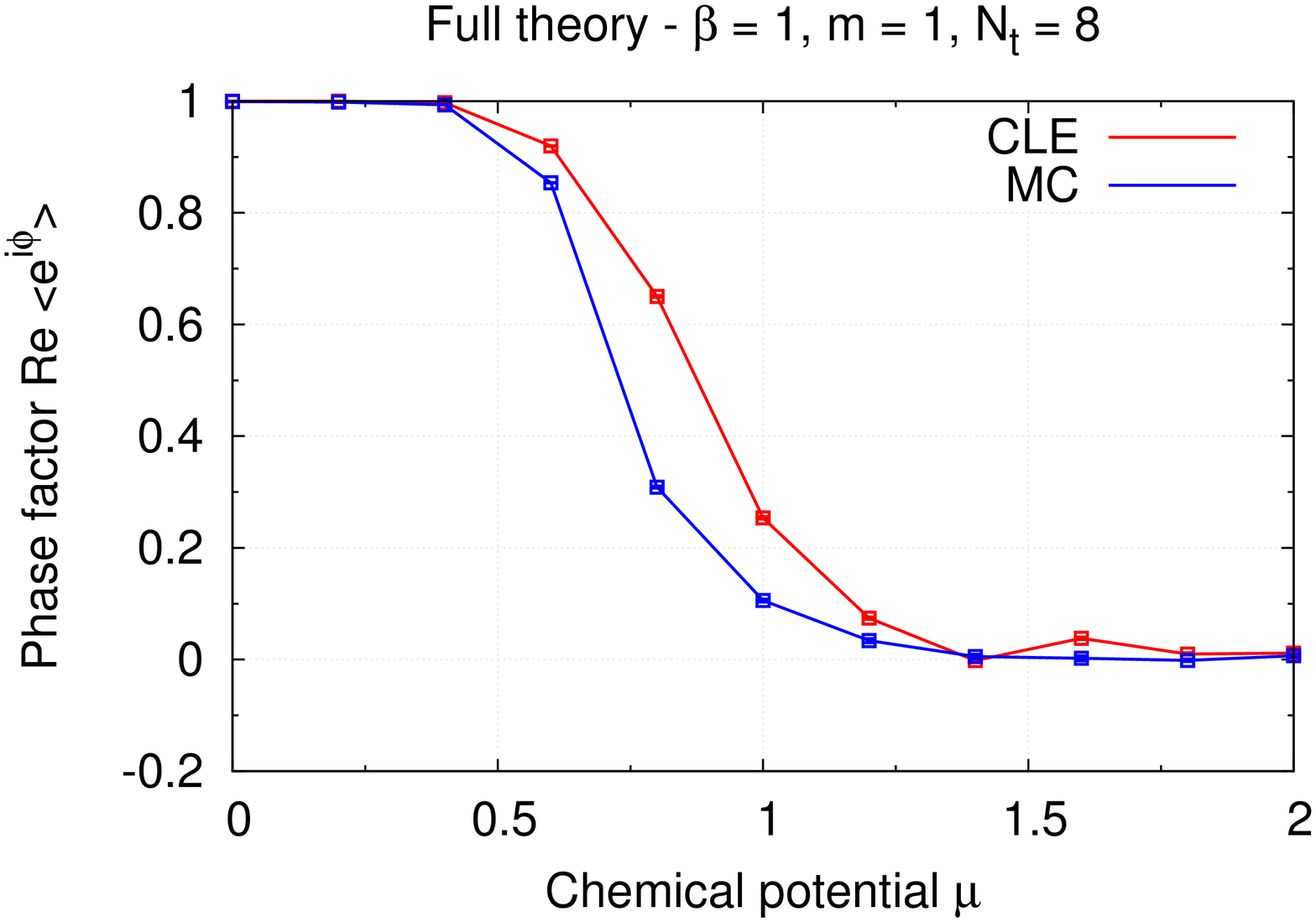}

}

\caption{Benchmarking results for $\beta=1$. \label{fig:Comparision-of-results-8-1-1}}
\end{figure}

We can see the results of a typical run for two different sets of
parameters in Figs.~\ref{fig:Comparision-of-results-4-3-1} and \ref{fig:Comparision-of-results-8-1-1}.
Because of the high sample size, the error bars tend to be very small.
We see that the results obtained with an adaptive Monte-Carlo method
are in good agreement with analytical results.

The numerical results obtained with a complex Langevin evolution show
some deviations for intermediate $\mu$. We systematically observe
that the gap widens for decreasing values of $\beta$, i.e., stronger
couplings. In addition the numerical evaluation becomes more noisy
as the relative magnitude of the noise term $\eta_{t}$ increases.
For large $\beta$ the agreement becomes better and the numerics are
very stable.

From the phase factor of the fermion determinant, we see how the sign
problem gets more severe for increasing lattice sizes. However, our
approach is not affected by this. It is interesting to note that the
sign problem seems to be less pronounced when using complex Langevin
evolutions.

In Fig.~\ref{fig:Comparision-of-results-8-1-1} we can already see
how in the limit $T\to0$ the Silver Blaze behavior becomes apparent.
In the limit $N_{t}\to\infty$ the observables will eventually become
independent of the chemical potential $\mu$ up to some threshold
$\mu_{c}$. It seems that the position of this onset differs slightly
from analytical predictions. However, for large $\beta$ they are
in good agreement.

\begin{figure}
\subfloat[Fermion density $\left\langle n\right\rangle $.]{\includegraphics[width=1\columnwidth]{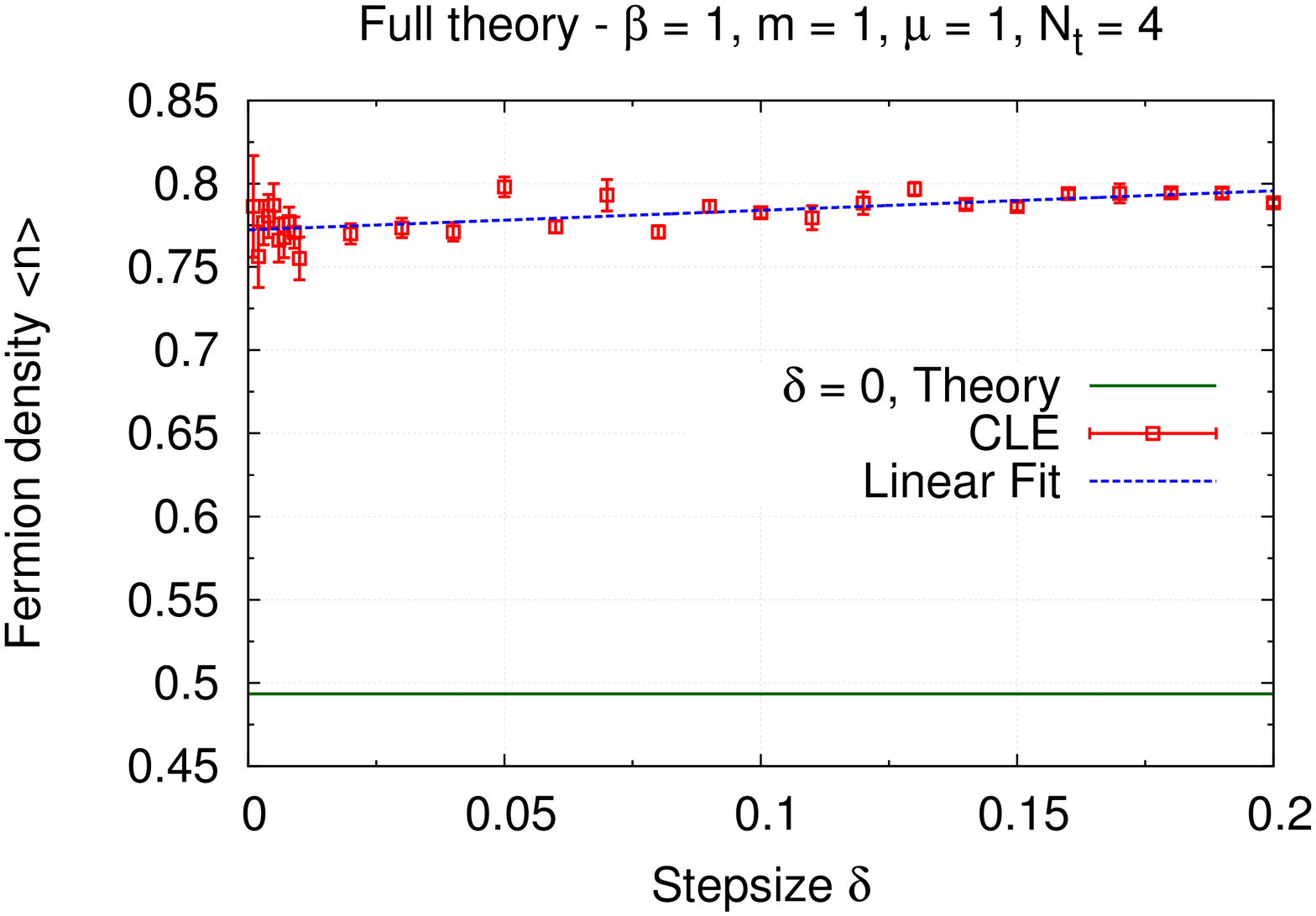}

}

\subfloat[Fermion condensate $\left\langle \overline{\chi}\chi\right\rangle $.]{\includegraphics[width=1\columnwidth]{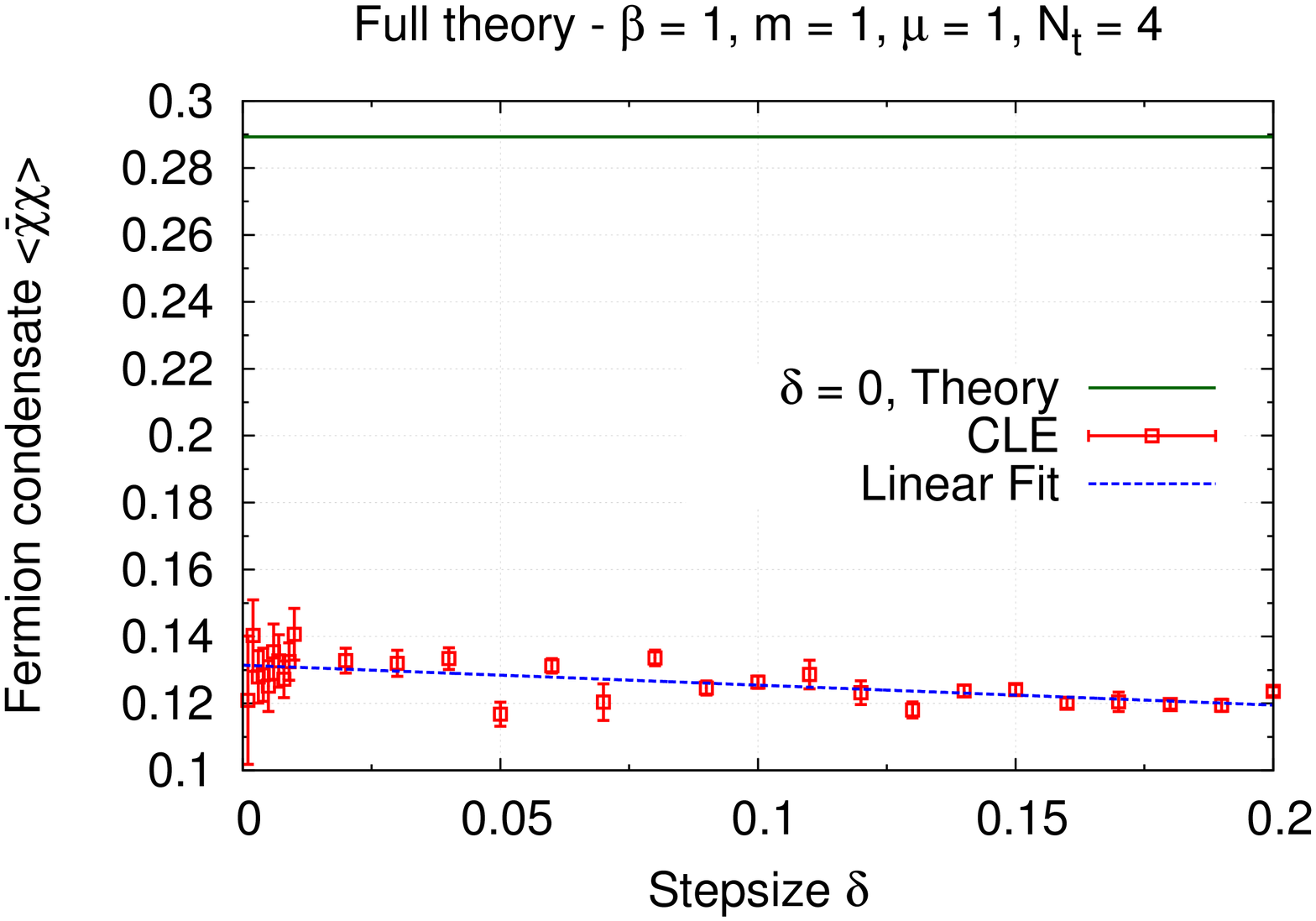}

}

\caption{Extrapolation to vanishing stepsize $\delta\to0$. \label{fig:Stepsize-Extrapolation}}
\end{figure}

That the observed deviations are not just an effect of a finite stepsize
can be seen in Fig.~\ref{fig:Stepsize-Extrapolation}. For a typical
set of parameters we calculate the fermion density and condensate
for varying stepsize factors $\delta$. The extrapolation shows that
also in the limit $\delta\to0$ there is a discrepancy between theory
and complex Langevin evolutions.

\begin{figure}
\includegraphics[width=1\columnwidth]{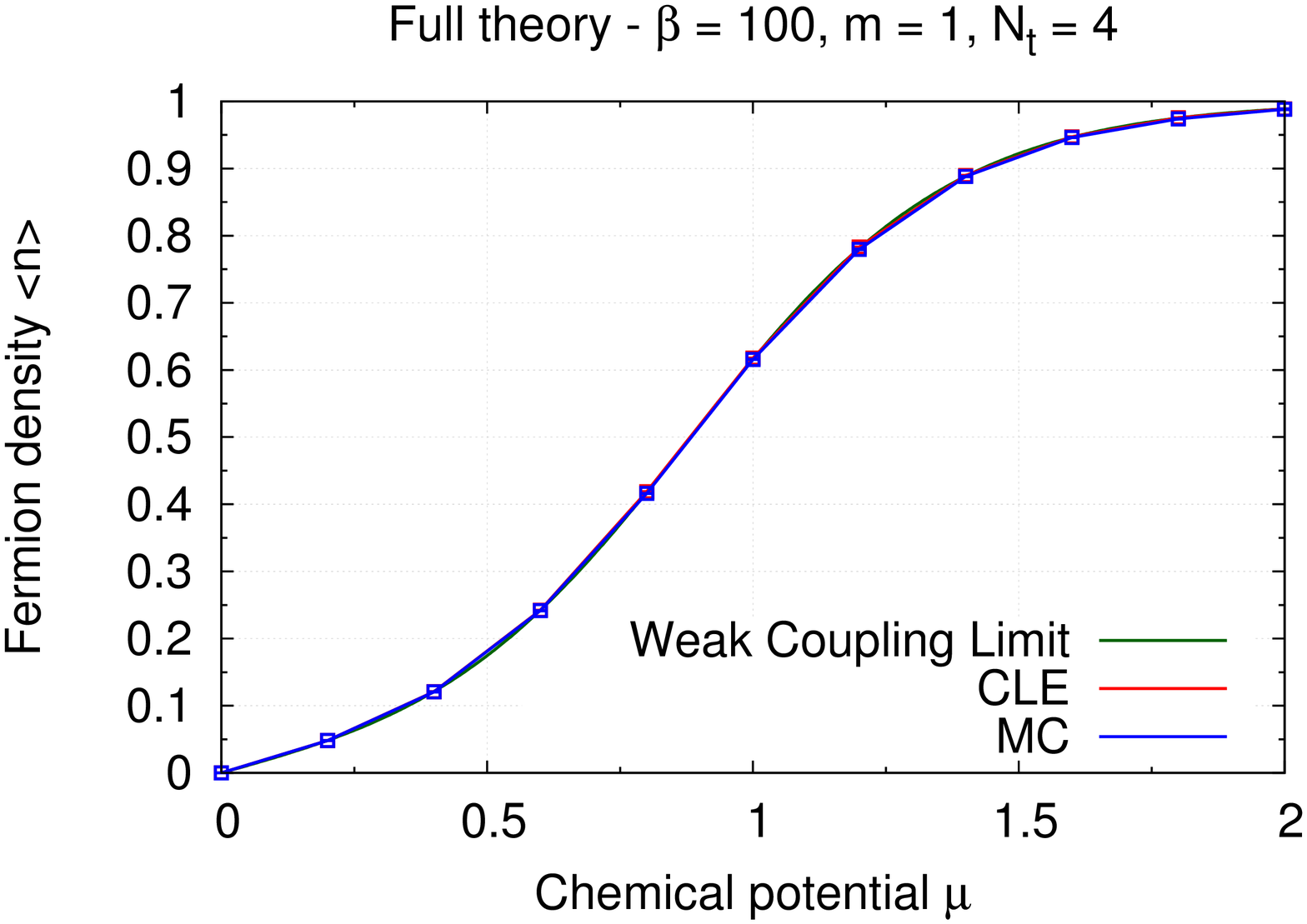}

\caption{Density in the weak coupling regime. \label{fig:Weak-Coupling}}
\end{figure}

\subsection{Coupling parameter dependence}

\begin{figure}
\subfloat[Fermion density $\left\langle n\right\rangle $.]{\includegraphics[width=1\columnwidth]{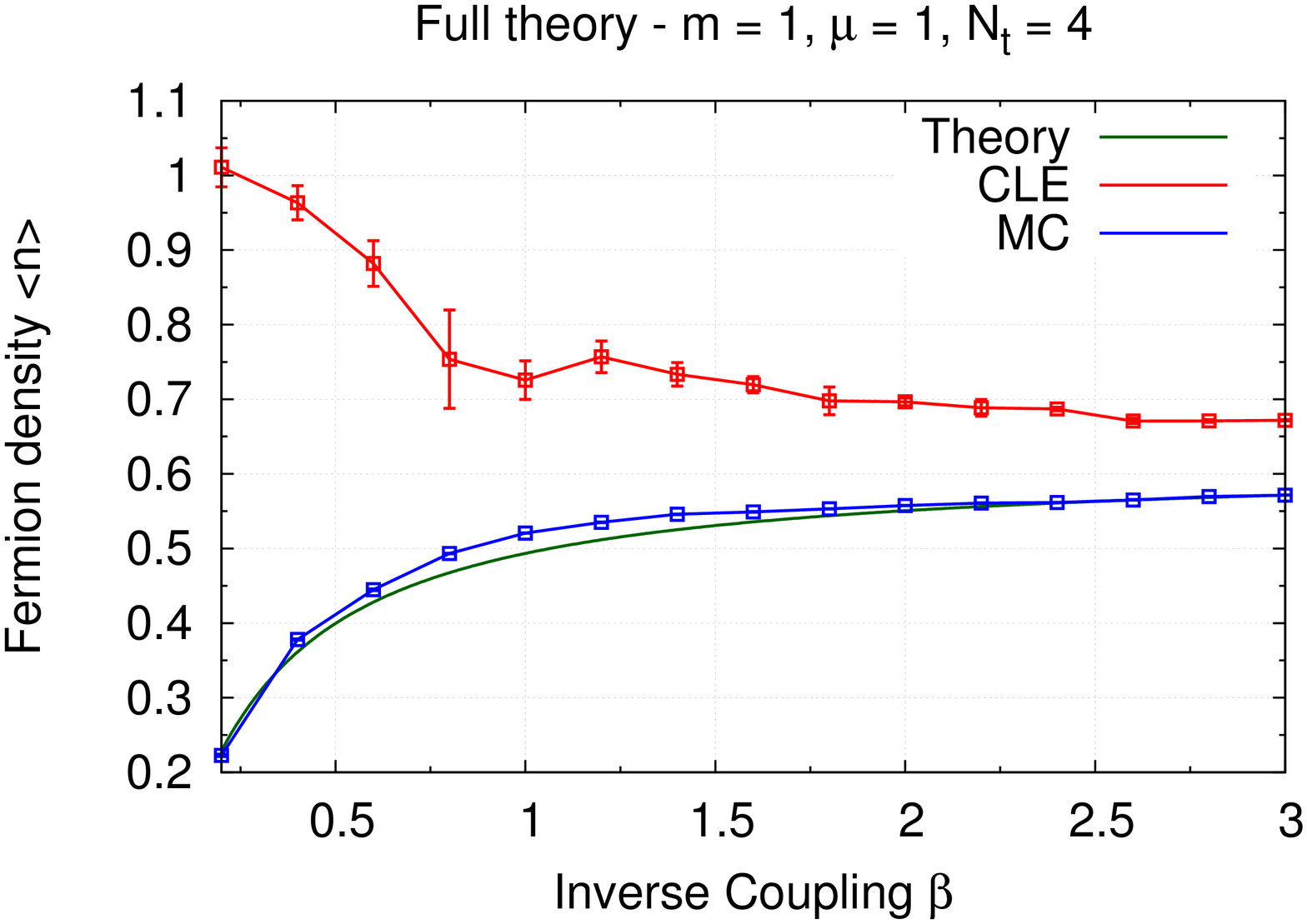}

}

\subfloat[Fermion condensate $\left\langle \overline{\chi}\chi\right\rangle $.]{\includegraphics[width=1\columnwidth]{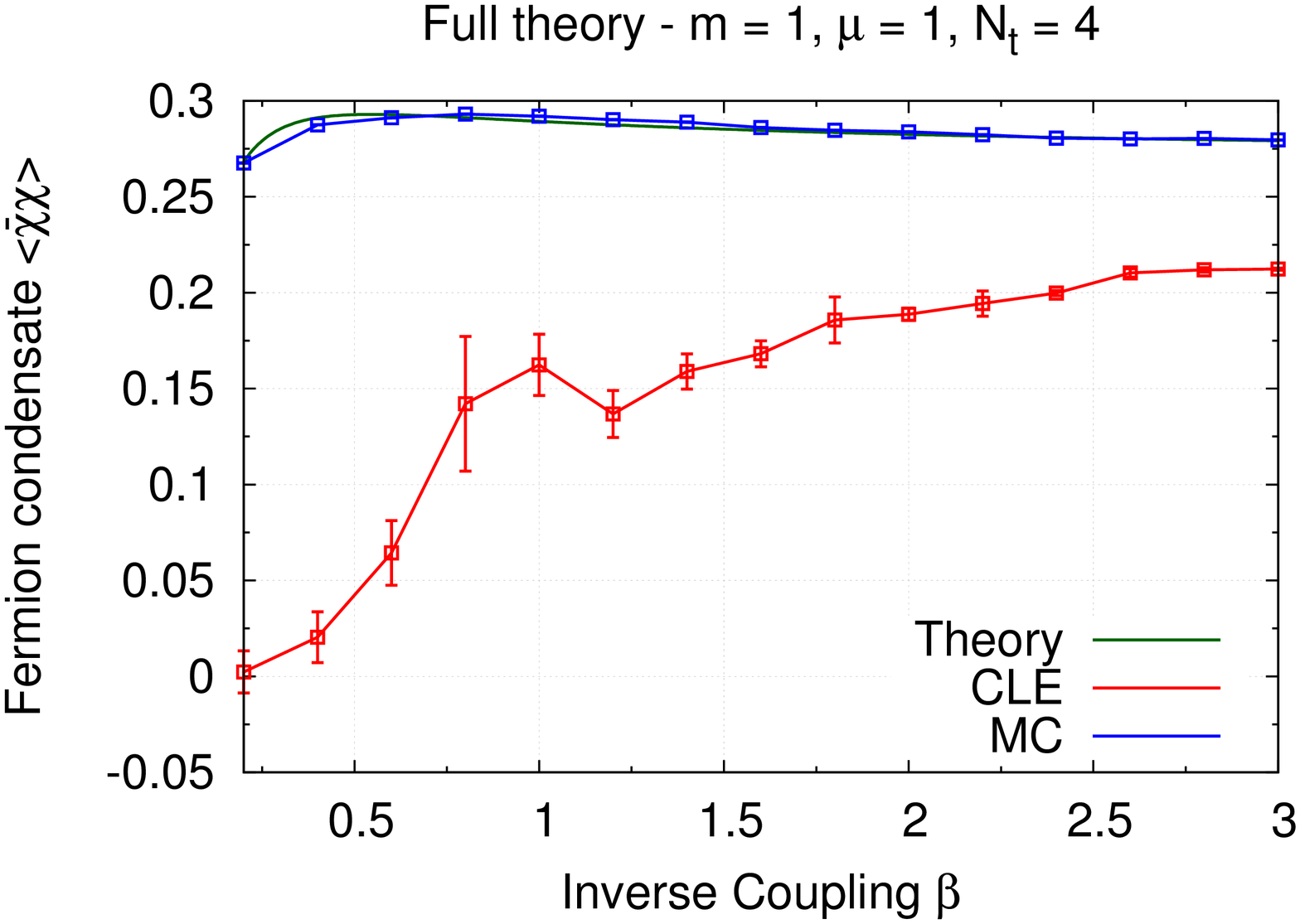}

}

\caption{Observables as function of $\beta$. \label{fig:Beta-Dependence}}
\end{figure}

In the weak coupling limit of Sec.~\ref{sub:Weak-coupling-limit},
we observe very good agreement of all numerical and analytical results.
For increasing $\beta$ the numerical results are asymptotically approaching
analytical results. In Figure \ref{fig:Weak-Coupling} we see that
for $\beta\gtrsim\mathcal{O}\left(100\right)$ the method is then
able to deliver reliable results. In the XY model at finite density
\cite{Aarts:2010vk} a similar behavior was observed. For small $\beta$
the authors observed incorrect convergence, while for large $\beta$
they found agreement. However, in the XY model there is a clear separation
between both regimes.

As already pointed out, the applicability of complex Langevin evolutions
depend on the magnitude of $\beta$. In Figure \ref{fig:Beta-Dependence}
we find the density and condensate as a function of $\beta$. We observe
that for weak couplings, i.e., in the limit $\beta\to\infty$, it
slowly converges towards the analytical results. For $\beta=\mathcal{O}\left(100\right)$
and above we then find good agreement. Contrariwise for strong couplings
we observe a large discrepancy. Moreover for $\beta\lesssim1$ the
evaluation of observables is very noisy, resulting in large error
bars.

\subsection{Several flavors \label{sub:Multi-flavor-numerical}}

\begin{figure}
\subfloat[Fermion density $\left\langle n\right\rangle $.]{\includegraphics[width=1\columnwidth]{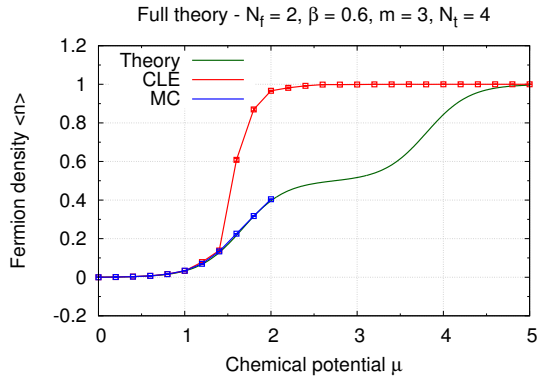}

}

\subfloat[Fermion condensate $\left\langle \overline{\chi}\chi\right\rangle $.]{\includegraphics[width=1\columnwidth]{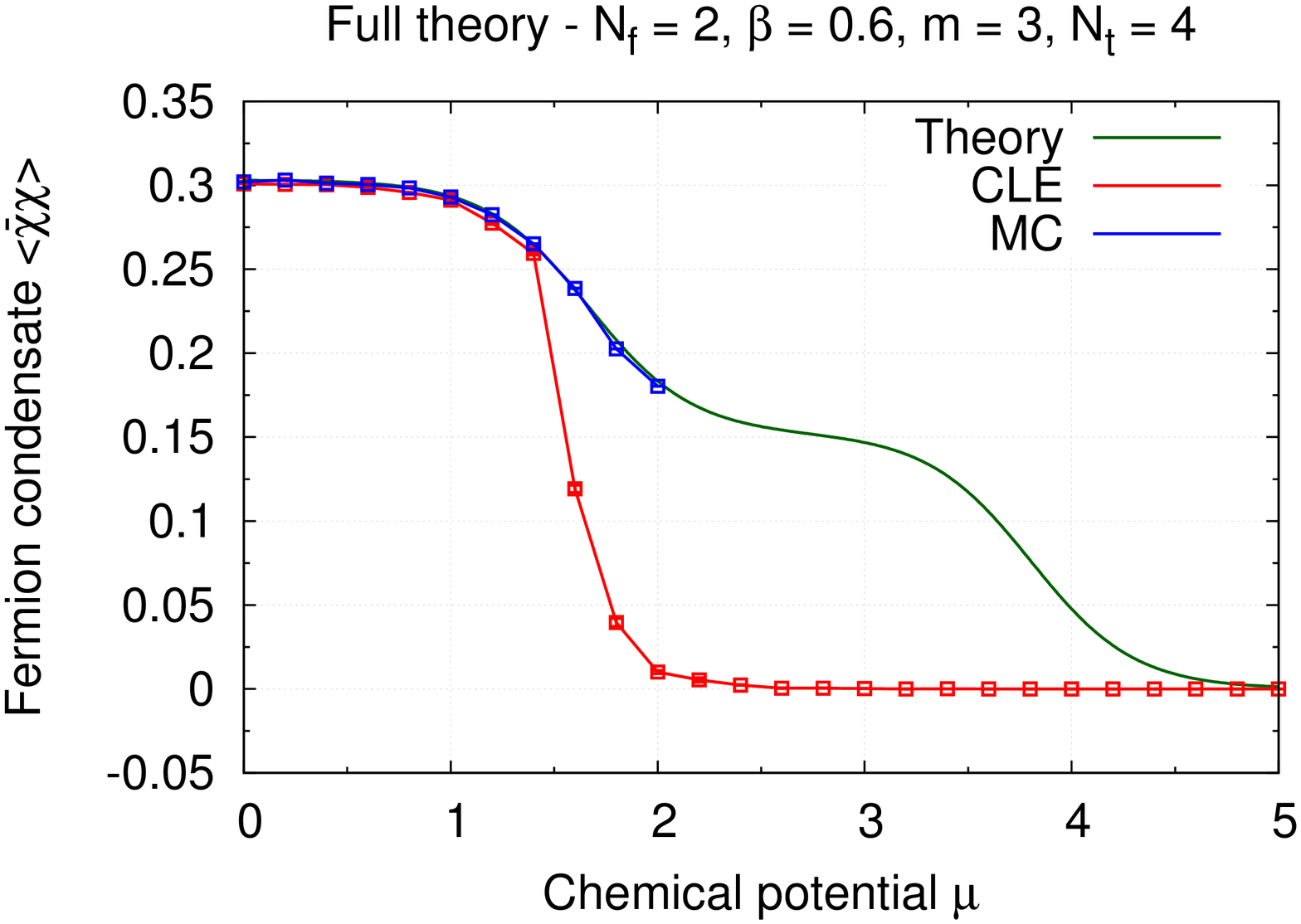}

}

\caption{Observables for $\mathcal{N}=2$ degenerated flavors. \label{fig:Multiflavor}}
\end{figure}

In Sec.~\ref{sub:Several-flavor-theory} we found that in a certain
range of $\beta$, the density, the condensate and the energy density
show up to $\mathcal{N}-1$ intermediate plateaus when considering
$\mathcal{N}$ flavors.

We give an example for $\mathcal{N}=2$ flavors at a coupling of $\beta=0.6$
in Fig.~\ref{fig:Multiflavor}. For small lattice sizes and small
$\mu$ Monte Carlo studies confirm these predictions. Because of the
sign problem we are restricted to $\mu\lesssim2$ with these particular
parameters, before the numerical evaluation breaks down.

When trying to reproduce the plateaus with a complex Langevin evolution,
we notice that for every value of $\beta$ the result qualitatively
resembles the $\mathcal{N}=1$ case. Directly after the onset the
fermion density rises until it reaches saturation. Only in the limit
of large $\beta$ the plateaus disappear in the analytical solution
and there is agreement with numerical results.

\subsection{Analyticity at $\mu^{2}=0$}

\begin{figure}
\includegraphics[width=1\columnwidth]{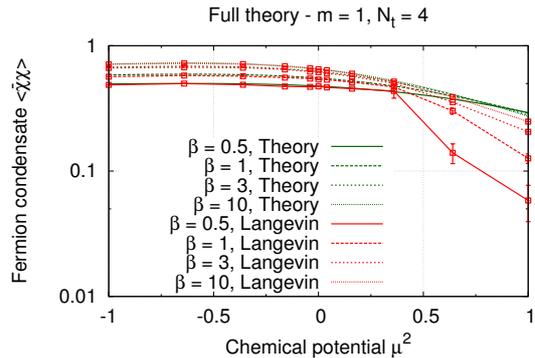}

\caption{Logarithmic plot of $\left\langle \overline{\chi}\chi\right\rangle $
over $\mu^{2}$. \label{fig:Condensate-in-mu2}}
\end{figure}

As described in Sec.~\ref{sub:Analiticity-in-mu2}, we check for
possible nonanalytic behavior of the condensate $\left\langle \overline{\chi}\chi\right\rangle $
at $\mu^{2}=0$. In Fig.~\ref{fig:Condensate-in-mu2} we see that
for $\mu^{2}\leq0$ the real Langevin evolution is, as expected, in
agreement with theory. Because of the absence of a sign problem numerical
results are reliable in this regime. Also for small $\mu^{2}\gtrsim0$
there is no statistically significant deviation. Hence the condensate
$\left\langle \overline{\chi}\chi\right\rangle $ is analytic within
our numerical accuracy.

If we increase $\mu$ further, we observe that at some point a disagreement
becomes apparent. The resulting gap is more pronounced for small $\beta$.
It does not appear suddenly, but develops smoothly and is visible
for all nonlarge $\beta$.

\subsection{Consistency conditions}

\begin{figure}[p]
\subfloat[Histogram of $\left\langle Ln\right\rangle $.]{\includegraphics[width=1\columnwidth]{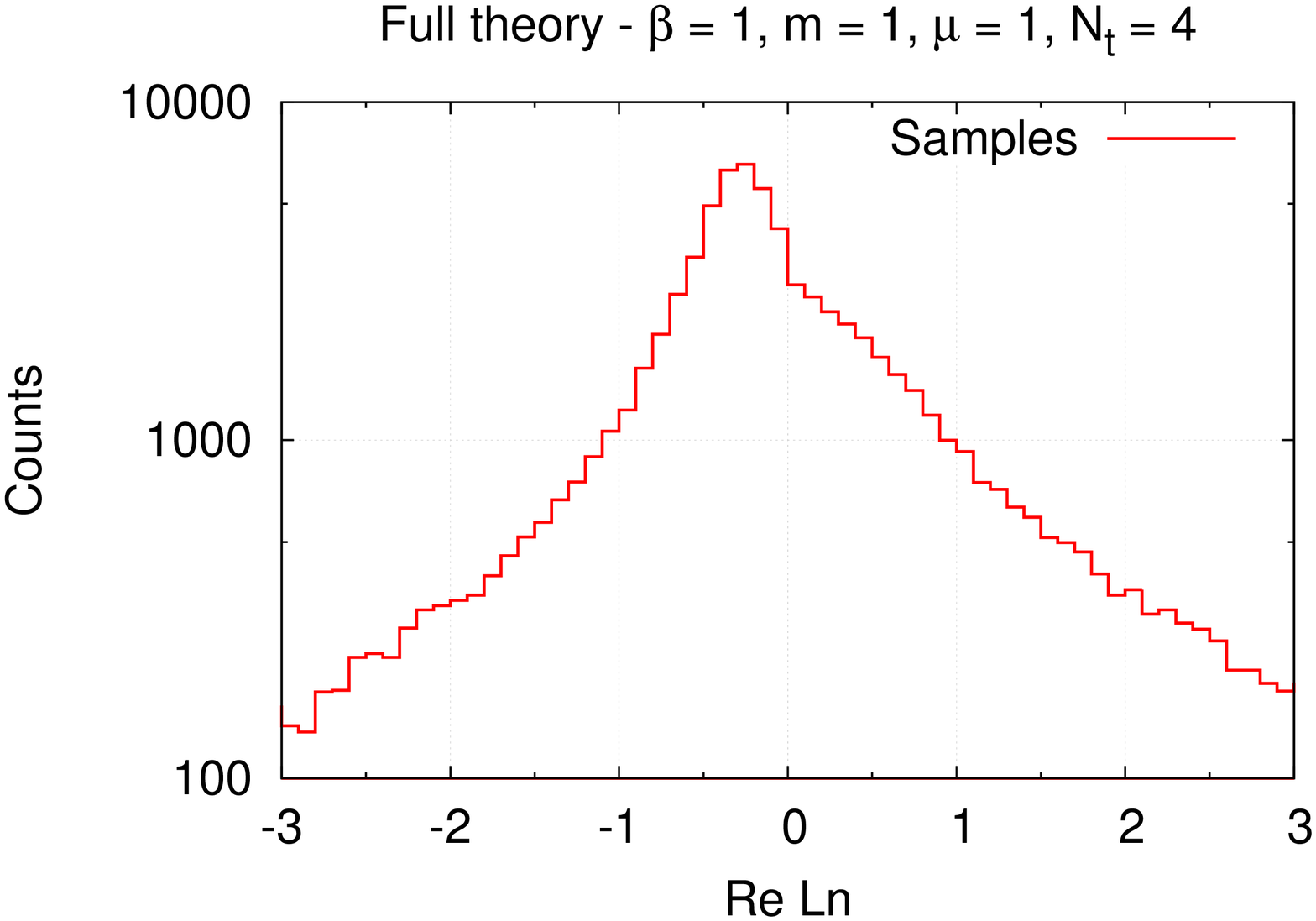}

}

\subfloat[Histogram of $\left\langle L\overline{\chi}\chi\right\rangle $.]{\includegraphics[width=1\columnwidth]{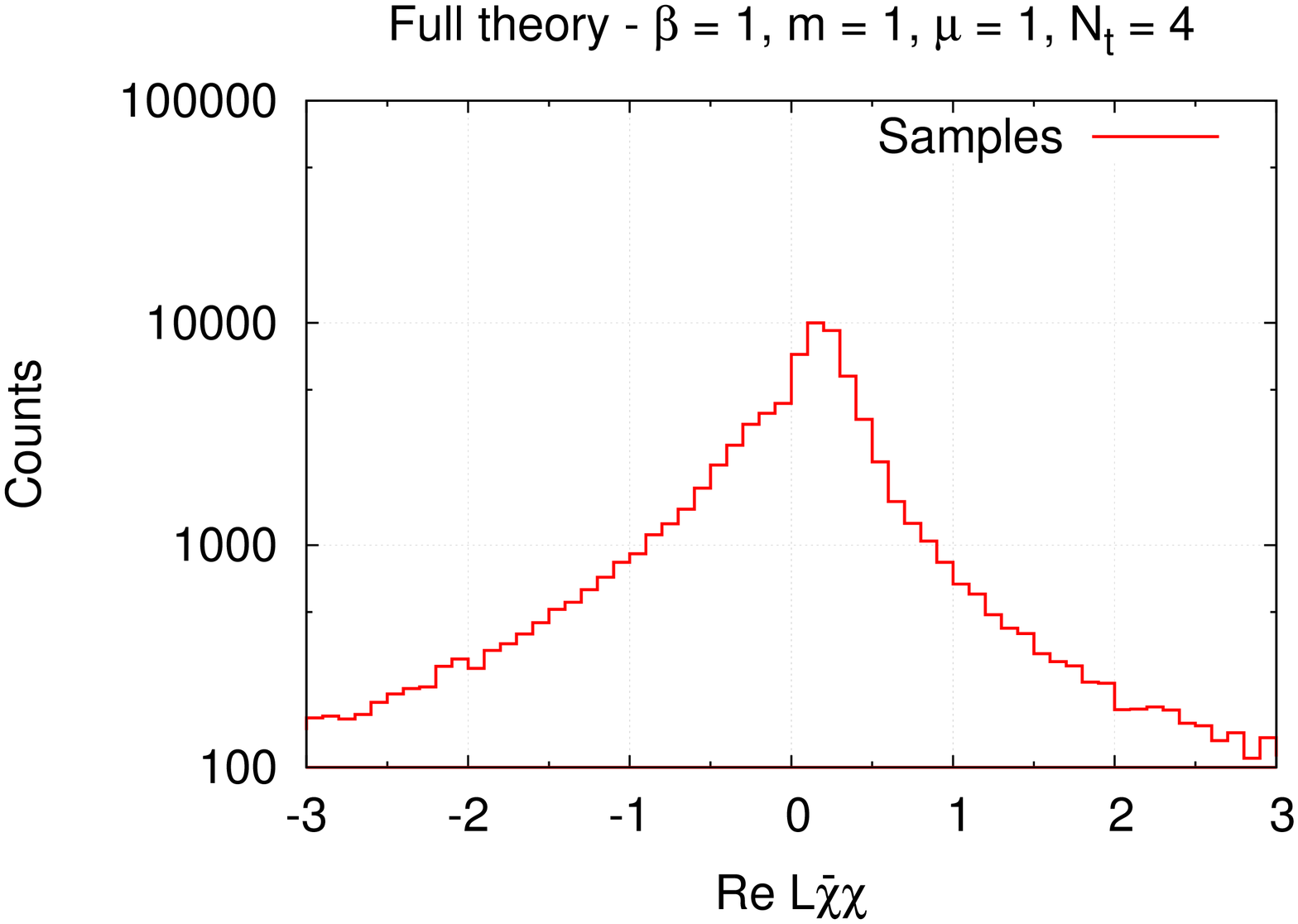}

}

\subfloat[Histogram of $\left\langle L\mathcal{O}\left(1,1\right)\right\rangle $.]{\includegraphics[width=1\columnwidth]{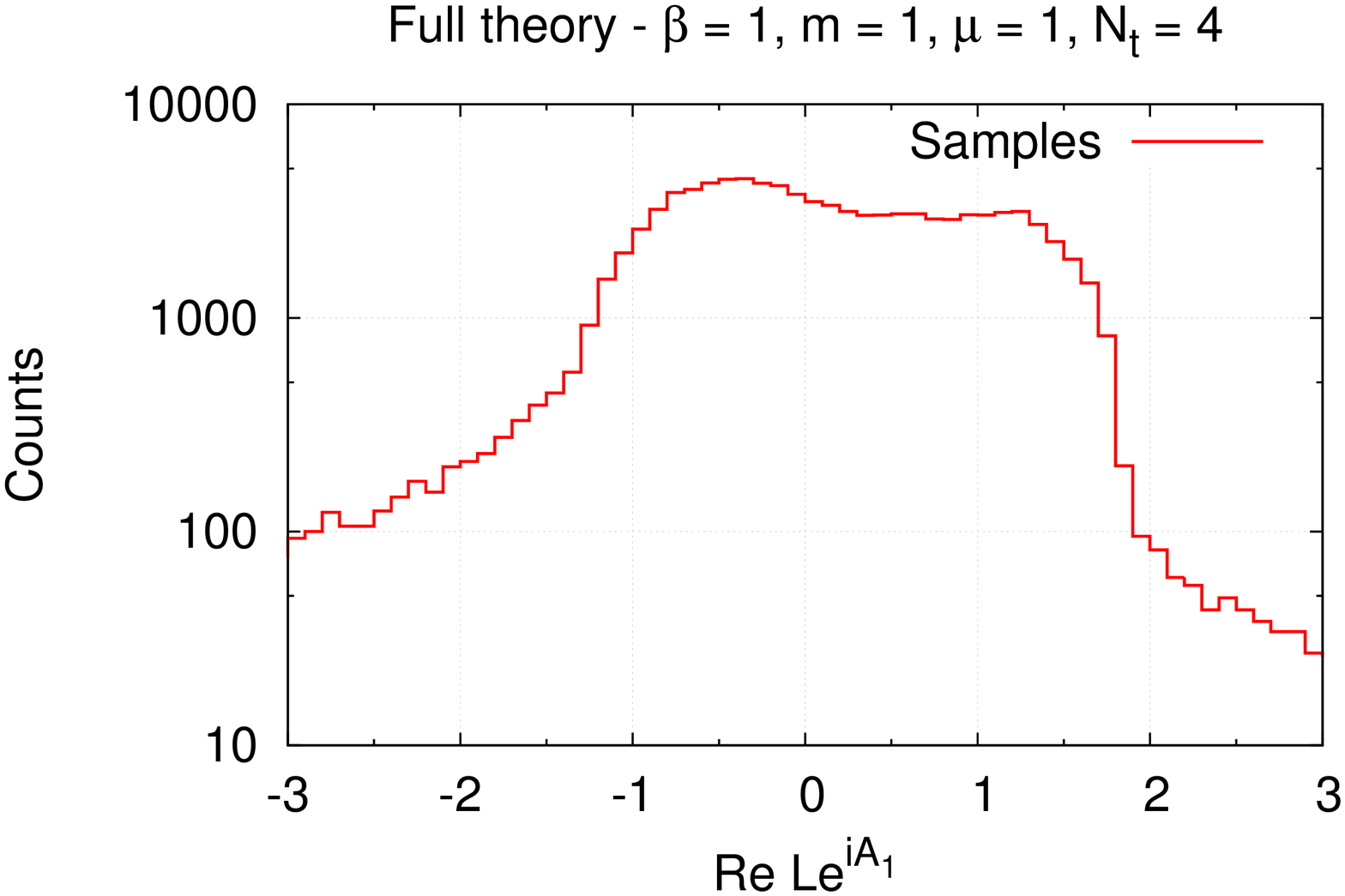}

}

\caption{Histograms for different consistency conditions. \label{fig:Histogram-consistency-conditions-0+1}}
\end{figure}
\begin{figure}
\includegraphics[width=1\columnwidth]{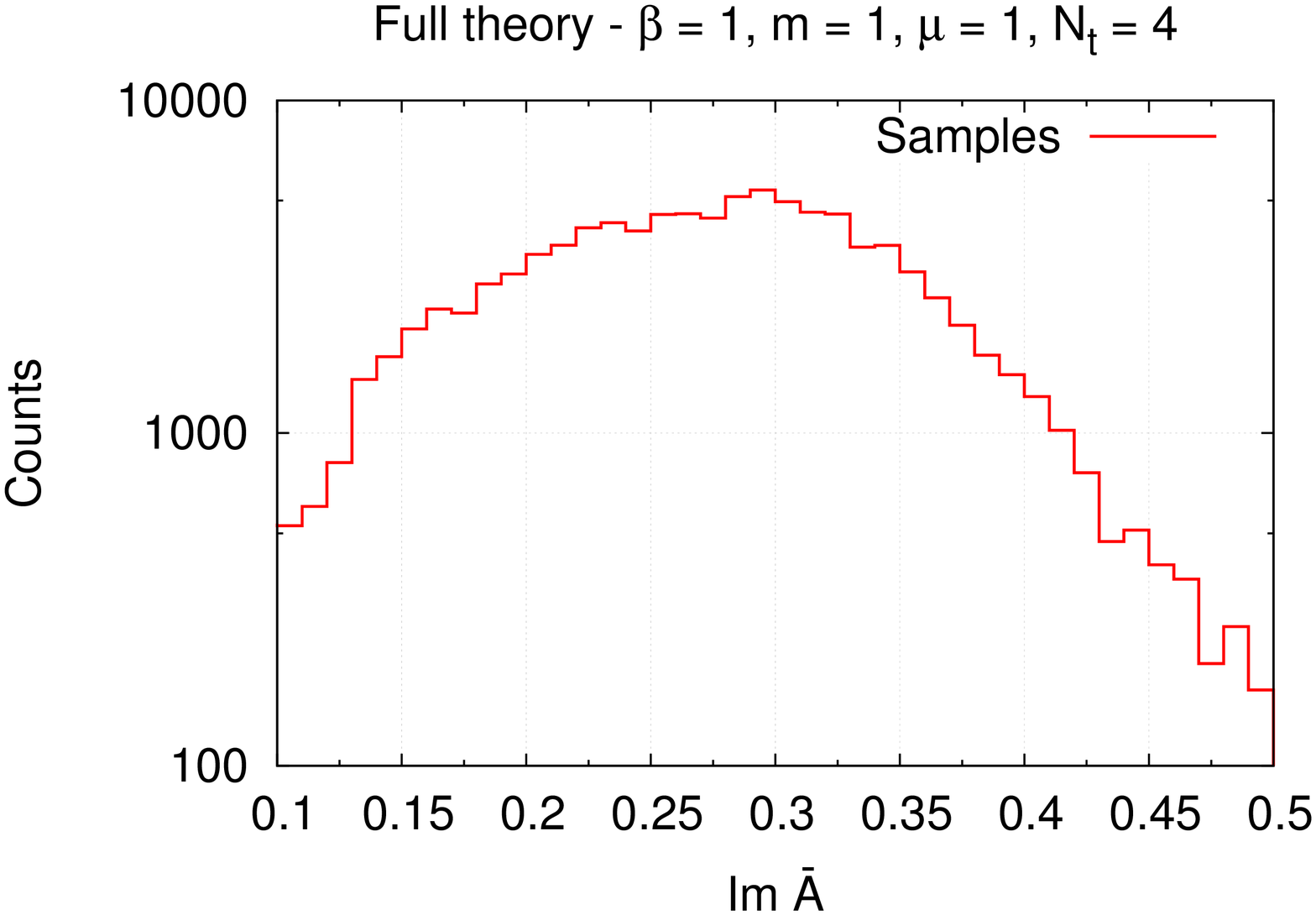}

\caption{Distribution of $\left\langle \myIm\overline{A}\right\rangle $. \label{fig:Histogram-field-0+1}}
\end{figure}

We checked the consistency conditions of Sec.~\ref{sub:Consistency-conditions-theory}
for the density $\left\langle n\right\rangle $, the condensate $\left\langle \overline{\chi}\chi\right\rangle $,
and $\mathcal{O}\left(t=1,k\right)$ for different sets of parameters
at $\delta=10^{-3}$. Before we begin with the interpretation of the
consistency conditions, we point out again the problem of using the
standard deviation to estimate error bounds. The resulting error is
much larger than the actual observed statistical fluctuations and
typically we obtain results like e.g.~$\myRe\left\langle L\mathcal{O}\left(1,1\right)\right\rangle =0.0152\pm6.1221$
for $N_{t}=4$, $\mathcal{I}=0$, $\mathcal{N}=\beta=m=\mu=1$. Here
$N_{t}$ denotes the temporal extension of the lattice, $\mathcal{N}$
the number of degenerated flavors, $\mathcal{I}$ the imaginary noise,
$\beta$ the inverse coupling constant, $m$ the mass and $\mu$ the
chemical potential. The overestimation of the error makes a meaningful
interpretation of the conditions impossible. Hence we estimate the
error with a bootstrap analysis.

We begin with the consistency conditions of the density and the condensate.
The evaluation is extremely noisy and only becomes slightly more stable
for large values of $\beta$. Without loss of generality we restrict
ourselves to the case of $t=1$. As an example we quote $\myRe\left\langle Ln\right\rangle =\left(143\pm305\right)\times10^{3}$
and $\myRe\left\langle L\overline{\chi}\chi\right\rangle =\left(-52\pm167\right)\times10^{3}$
for $N_{t}=4$, $\mathcal{I}=0$, $\mathcal{N}=\beta=m=\mu=k=1$.
The resulting expressions seem to be numerically ill-defined and despite
large sample sizes, we are unable to draw any conclusions about the
validity of the conditions. In our case both observables proved to
be unsuitable to check for the correctness of the complex Langevin
evolution.

The evaluation of the conditions for the observable $\left\langle \mathcal{O}\left(t,k\right)\right\rangle $
on the other hand is stable for all checked parameter sets. If we
take the error bounds seriously, we have to conclude that some conditions
are not compatible with a vanishing value and are violated. We quote
here $\myRe\left\langle L\mathcal{O}\left(t,k\right)\right\rangle =-0.1395\pm0.0110$
for $N_{t}=4$, $\mathcal{I}=0$, $\mathcal{N}=\beta=m=k=1$, $\mu=3$
as a typical example of a violated condition and $\myRe\left\langle L\mathcal{O}\left(t,k\right)\right\rangle =0.0061\pm0.0208$
for $N_{t}=4$, $\mathcal{I}=0$, $\mathcal{N}=\beta=m=\mu=k=1$ as
one which is compatible with a vanishing value. We also checked the
conditions for $k=2,3$, which turned out to be more noisy compared
to the $k=1$ case. We interpret the violated conditions as an indicator
for incorrect convergence of the complex Langevin evolution.

When considering the distributions of the evaluated consistency conditions,
we can gain further insights. In Fig.~\ref{fig:Histogram-consistency-conditions-0+1}
we find histograms of $Ln$, $L\overline{\chi}\chi$, and $L\mathcal{O}\left(1,1\right)$.
In general the distributions are asymmetrical and non-Gaussian. In
the case of the density and condensate we saw that the distribution
falls off extremely slowly and we observed values with an absolute
value of up to $\mathcal{O}\left(10^{9}\right)$, resulting in a very
noisy evaluation. In contrast the distribution of $L\mathcal{O}\left(1,1\right)$
falls off rapidly.

It is also interesting to check the distribution of the imaginary
part of the auxiliary field itself. In the derivation of the consistency
condition, one assumes that boundary terms of the field would vanish.
It is then important to check that the imaginary part falls off rapidly
enough. In Fig.~\ref{fig:Histogram-field-0+1} we find a typical
histogram of the average imaginary part $\myIm\overline{A}=N_{t}^{-1}\sum_{t}\myIm A_{t}$
with $10^{5}$ samples. We see that the resulting distribution appears
to be compatible with our hypothesis.

\subsection{Eigenvalues of the fermion matrix}

\begin{figure}
\includegraphics[clip,width=1\columnwidth]{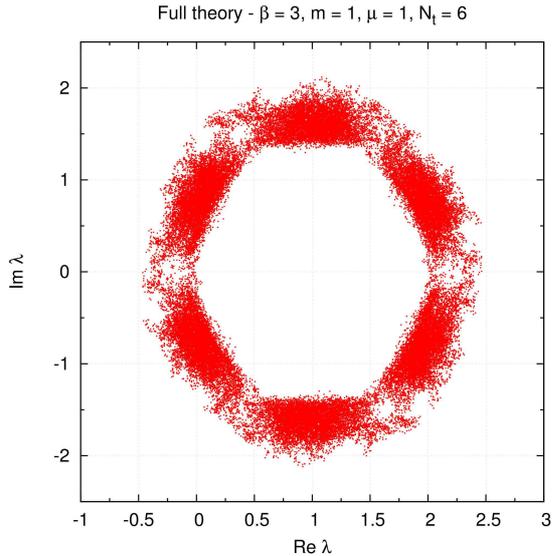}

\caption{Eigenvalues of the fermion matrix. \label{fig:Scatterplot-eigenvalues}}
\end{figure}

In Figure \ref{fig:Scatterplot-eigenvalues} we find a scatter plot
of the eigenvalues of the fermion matrix $K$ defined in \eqref{eq:Fermion-matrix}.
We sampled approximately $6\times10^{4}$ eigenvalues during a complex
Langevin evolution. Eigenvalues come in point-reflected pairs at point
$\left(m,0\right)$, where $m$ denotes the mass of the fermion. For
a vanishing chemical potential $\mu=0$ all eigenvalues lie on the
line $\myRe\lambda=m$.

\subsection{Imaginary noise}

\begin{figure}
\includegraphics[width=1\columnwidth]{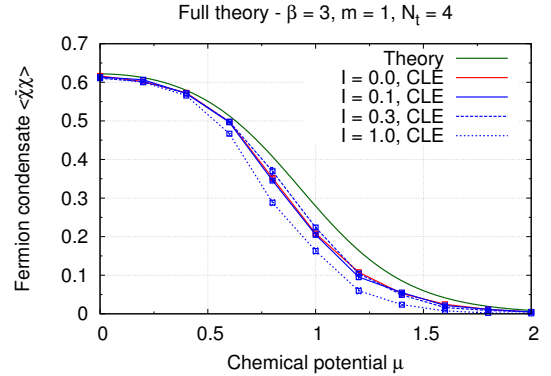}

\caption{Condensate $\left\langle \overline{\chi}\chi\right\rangle $ with
imaginary noise $\mathcal{I}$. \label{fig:Imaginary-Noise}}
\end{figure}

Assuming correct convergence of the complex Langevin evolution, observables
should turn out independent of the imaginary noise term controlled
by $\mathcal{I}$. However, in Fig.~\ref{fig:Imaginary-Noise} we
actually observe such a dependence. This was observed previously in
other systems too \cite{Aarts:2011ax}. Attempts to fine-tune $\mathcal{I}$
so that the complex Langevin evolution is deformed and correctly reproduces
analytical results did not succeed. In almost all cases an imaginary
noise $\mathcal{I}>0$ caused a more severe disagreement between numerical
and analytical results.

\subsection{Periodic continuation of the imaginary part}

\begin{figure}
\includegraphics[width=1\columnwidth]{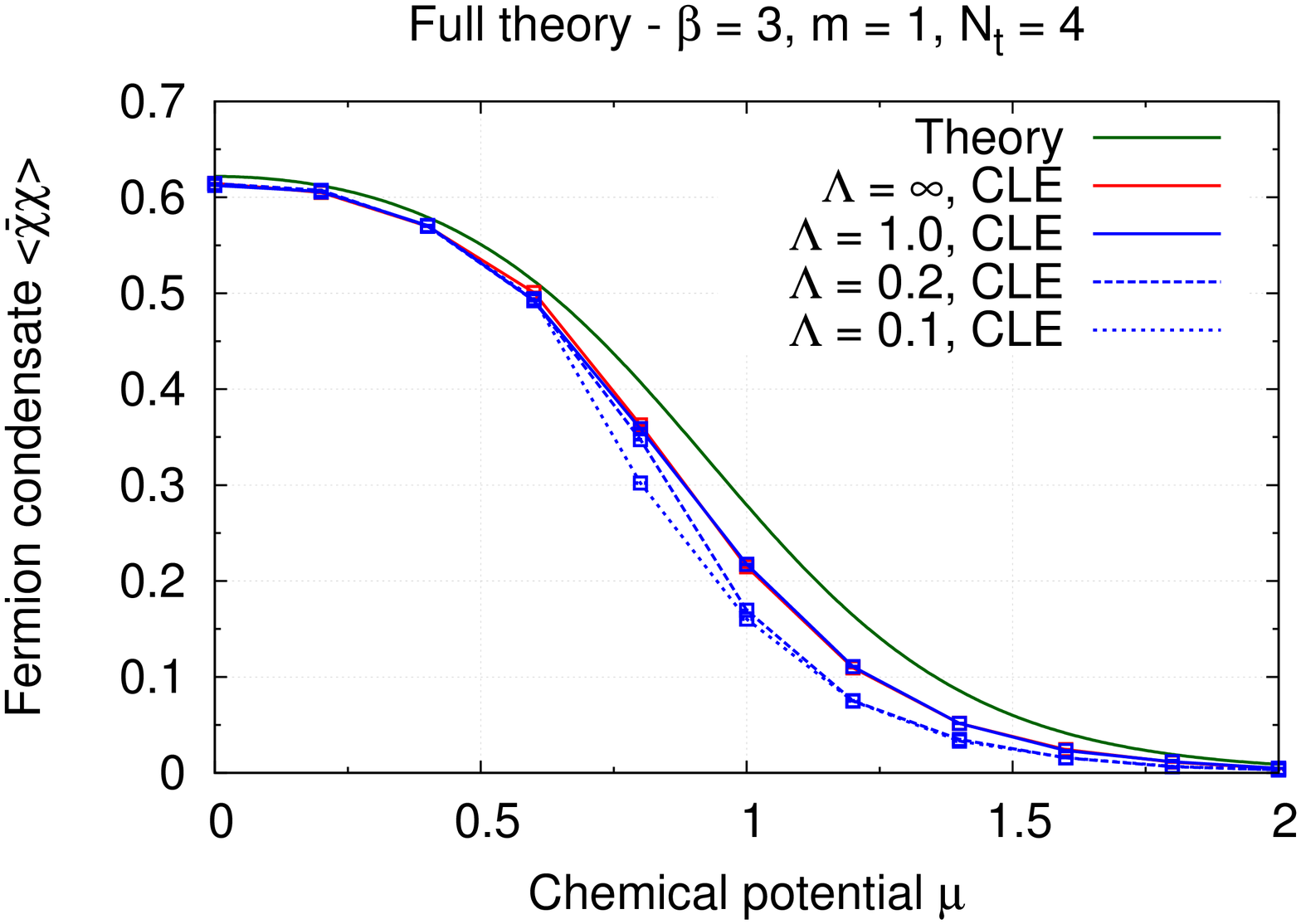}

\caption{Condensate $\left\langle \overline{\chi}\chi\right\rangle $ with
periodic cutoff $\Lambda$. \label{fig:Periodic-Im-Part}}
\end{figure}

A necessary condition for the correctness of the complex Langevin
evolution is that the distributions of the imaginary parts of the
fields have to fall off rapidly enough. Although in the generalized
Thirring model this seems to be the case, we can restrict the imaginary
part of the field to the interval $\left[-\Lambda,+\Lambda\right]$
and make a periodic continuation. A similar approach was employed
in \cite{Aarts:2011ax} (see also \cite{Aarts:2009uq}), where a fine-tuning
of the cutoff enabled simple toy models to reproduce correct results.
However, in Fig.~\ref{fig:Periodic-Im-Part} we see that every finite
value of $\Lambda$ widens the gap to the correct results. Hence in
this simple form this approach cannot be applied to the generalized
Thirring model.

\section{Conclusions \label{sec:Conclusions}}

In this paper we applied a complex Langevin evolution to a generalized
Thirring model in $0+1$ dimensions in order to deal with the resulting
sign problem for $\mu>0$. For intermediate values of the chemical
potential we found a gap between analytical and numerical results,
which size depends on the inverse coupling $\beta$. While for small
$\beta$ the discrepancy is large, we observe agreement for large
$\beta$. In particular, for $\beta\gtrsim\mathcal{O}\left(100\right)$
we are usually able to reproduce analytical results with high accuracy.
Furthermore for small $\mu\gtrsim0$ we did not observe any significant
deviations to theoretical predictions and the fermion condensate is
analytic at $\mu^{2}=0$.

However, in the case of more than one flavor we observed a qualitative
disagreement. Our approach seems to be unable to reproduce the plateaus
we found for certain ranges of $\beta$. Another interesting observation
is the violation of several consistency conditions, indicating that
the complex Langevin evolution might not converge correctly in general.
Attempts to force correct convergence by an \textit{ad hoc} fine-tuning
of a periodic cutoff $\Lambda$ or an imaginary noise term $\mathcal{I}$
did not succeed.

In a subsequent paper, we will present our findings for the generalized
Thirring model in $2+1$ dimensions \cite{Pawlowski:2013gag}. Further
investigations have to deal with the question of how to address the
aforementioned problems. In particular, coordinate transformations
as suggested in \cite{Aarts:2012ft} and gauge cooling procedures
like the one employed in \cite{Seiler:2012wz} might allow a stabilization
of the complex Langevin evolution.

\section*{Acknowledgments}

We thank I.-O.~Stamatescu for uncountable discussions and useful
remarks on the manuscript. We also thank C.~Gattringer for a proof
of \eqref{eq:Latt-Exact-Part-Func} by systematic saturation of the
Grassmann integral and V.~Kasper for a proof using the determinant
identities in \cite{Molinari20082221}. Furthermore we acknowledge
E.~Seiler's derivation of the plateau structures as presented in
the Appendix. Finally, we thank G.~Aarts and D.~Sexty for discussions.
This work is supported by the Helmholtz Alliance HA216/EMMI and by
ERC-AdG-290623. C.~Z.~thanks the German National Academic Foundation
for financial support.\bigskip{}

\appendix

\section*{Appendix: Plateaus \label{sec:App-plateaus}}

\begin{figure}
\includegraphics[width=1\columnwidth]{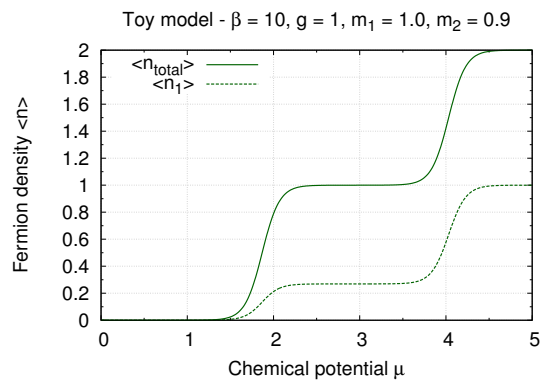}

\caption{Densities in the toy model. \label{fig:Toy-model-plateaus-FD}}
\end{figure}

As previously discussed, in the case of $\mathcal{N}>1$ flavors we
observe intermediate plateaus in the considered observables. In the
nonlinear O(2) sigma model, the authors of \cite{Banerjee:2010kc}
found similar structures. They explained this finite size behavior
with the crossing of energy levels at finite chemical potential. However,
here we reproduce them with the help of a continuum model, which corresponds
to the $0+1$ dimensional generalized Thirring model. To this end
we consider the Hamiltonian
\begin{equation}
H=\sum_{i=1}^{N_{f}}m_{i}\left(n_{i}+\overline{n}_{i}\right)+g^{2}\left(\sum_{i=1}^{N_{f}}Q_{i}\right)^{2}
\end{equation}
with flavor index $i=1,\dots,N_{f}$ and bare masses $m_{i}$. We
introduced $Q_{i}\equiv n_{i}-\overline{n}_{i}$, the particle number
operator $n_{i}=a_{i}^{\dagger}a_{i}$ and the antiparticle number
operator $\overline{n}_{i}=b_{i}^{\dagger}b_{i}$, which are given
in terms of ladder operators. From the grand canonical partition function
\begin{equation}
Z=\Tr\exp\left[-\beta_{T}\left(H-\sum_{i}\mu_{i}Q_{i}\right)\right]
\end{equation}
with $\beta_{T}=T^{-1}$ and temperature $T$, we can derive the fermion
densities
\begin{equation}
\left\langle n\right\rangle _{i}=\frac{1}{\beta_{T}}\frac{\partial\log Z}{\partial\mu_{i}},\quad\left\langle n\right\rangle _{\textrm{total}}=\sum_{i}\left\langle n\right\rangle _{i}.
\end{equation}
The plateaus are a result of the competition between the $Q_{i}$
and $Q_{i}^{2}$ terms. A typical plot for $N_{f}=2$ can be found
in Fig.~\ref{fig:Toy-model-plateaus-FD}. Like in the Thirring model
they appear for intermediate values of $g^{2}$ and eventually disappear
in the limit of small and large values.

\bibliography{literature}

\end{document}